\documentclass[apj]{emulateapj}
 \slugcomment{Received 2013 September 30;·
  accepted 2014 July 4}
\usepackage{psfig}
\usepackage{color}
\usepackage{hyperref}

\hypersetup{
  colorlinks=true,
  linkcolor=blue,
  citecolor=blue}

\newcommand{\gsim}{\mbox{\hspace{.2em}\raisebox{.5ex}{$>$}\hspace{-.8em}\raisebox{-.5ex}{$\sim$}\hspace{.2em}}}
\newcommand{\lsim}{\mbox{\hspace{.2em}\raisebox{.5ex}{$<$}\hspace{-.8em}\raisebox{-.5ex}{$\sim$}\hspace{.2em}}}
\newcommand{\ssst}{\scriptscriptstyle}
\newcommand{\E}[1]{\times 10^{#1}}
\newcommand{\etal}{et al.}

\newcommand{\s}{\,{\rm s}}      \newcommand{\ps}{\,{\rm s}^{-1}}
\newcommand{\yr}{\,{\rm yr}}    \newcommand{\Msun}{M_{\odot}}
\newcommand{\cm}{\,{\rm cm}}    \newcommand{\km}{\,{\rm km}}
\newcommand{\kms}{$\km\ps$}
\newcommand{\kpc}{\,{\rm kpc}} \newcommand{\pc}{\,{\rm pc}}
\newcommand{\erg}{\,{\rm erg}}        
    \newcommand{\keV}{\,{\rm keV}}

\newcommand{\nel}{n_{e}}        \newcommand{\NH}{N_{\ssst\rm H}}
\newcommand{\no}{n_{\ssst 0}}

\newcommand{\kTc}{kT_{\rm c}} \newcommand{\kTh}{kT_{\rm h}}
 \newcommand{\kTz}{kT_{\rm z}}
\newcommand{\Tc}{T_{\rm c}} \newcommand{\Th}{T_{\rm h}}
\newcommand{\tauc}{\tau_{\rm c}} \newcommand{\tauh}{\tau_{\rm h}}
         
\newcommand{\nH}{n_{\ssst\rm H}}        
\newcommand{\nHc}{n_{{\ssst\rm H,} c}} 
\newcommand{\nHh}{n_{{\ssst\rm H,} h}}

       \newcommand{\Einstein}{{\sl Einstein}}
\newcommand{\ROSAT}{{\sl ROSAT}} 

\newcommand{\XMMN}{{\sl XMM-Newton}}
\newcommand{\Chandra}{{\sl Chandra}}
\newcommand{\ASCA}{{\sl ASCA}}
\newcommand{\Suzaku}{{\sl Suzaku}}
\newcommand{\Fermi}{{\sl Fermi}}
\newcommand{\du}{d_{2}}

\newcommand{\rray}{$\gamma$-ray}
\newcommand{\Ha}{H$\alpha$}
\newcommand{\Hea}{He$\alpha$}
\newcommand{\Lya}{Ly$\alpha$}
\newcommand{\Ka}{K$\alpha$}
\newcommand{\SII}{\ion{S}{2}}

\newcommand{\snr}{W28}
\newcommand{\twCO}{$^{12}$CO}   
    
\newcommand{\Jttt}{$J$=3--2}

\newcommand{\vmekal}{$vmekal$}

\newcommand{\vnei}{$vnei$}
\newcommand{\neij}{$neij$}
\newcommand{\powerlaw}{$powerlaw$}

\newcommand{\HI}{\ion{H}{1}}

\newcommand{\rmxaa}{RMxAA}

\shorttitle{}

\begin{document}

\title{AN \XMMN\ STUDY OF THE MIXED-MORPHOLOGY SUPERNOVA REMNANT \snr\ 
(G6.4$-$0.1)}

\author{
 Ping Zhou\altaffilmark{1,2},
 Samar Safi-Harb\altaffilmark{2,4},
 Yang Chen\altaffilmark{1,3},
 Xiao Zhang\altaffilmark{1},
 Bing Jiang\altaffilmark{1},
 and Gilles Ferrand\altaffilmark{2}
}
\affil{$^1$School of Astronomy and Space Science, Nanjing University, 
Nanjing~210093, China \\
$^2$Department of Physics and Astronomy, University of
Manitoba, Winnipeg~R3T 2N2, Canada \\
$^3$ Key Laboratory of Modern Astronomy and Astrophysics,
Nanjing University, Ministry of Education, China}
\altaffiltext{4}{Canada Research Chair}

\begin{abstract}
We have performed an \XMMN\ imaging and spectroscopic study of supernova
remnant (SNR) \snr, a prototype mixed-morphology or thermal composite SNR
believed to be interacting with
a molecular cloud.
The observed hot X-ray emitting plasma is characterized by low
metal abundances, showing no evidence of ejecta.
The X-rays arising from the deformed northeast shell consist of a 
thermal component with a temperature of $\sim0.3$ keV plus a hard 
component of either thermal (temperature $\sim 0.6$~keV) 
or non-thermal (photon index $=0.9$--2.4) origin.
The X-ray emission in the SNR interior is blobby and the corresponding
spectra are best described as the emission from a cold ($kT$~$\sim$~0.4
keV) plasma in non-equilibrium ionization with an ionization 
timescale of $\sim4.3\E{11}$~$\cm^{-3}\s$ plus a hot 
($kT \sim0.8$~keV) gas in collisional ionization equilibrium.
Applying the two-temperature model to the smaller central regions,
we find non-uniform interstellar absorption, temperature and 
density distribution,
which indicates that the remnant is evolving in a non-uniform 
environment with denser material in the east and north.
The cloudlet evaporation mechanism can essentially explain the 
properties of the X-ray emission in the center
and thermal conduction may also play a role for length scales comparable 
to the remnant radius.
A recombining plasma model with an electron temperature of
$\sim 0.6$~keV is also feasible for describing the hot central gas
with the recombination age of the gas estimated at $\sim2.9\E{4}$~yr.

\end{abstract} 


\section{Introduction} \label{S:intro}
Mixed-morphology (MM) or thermal composite supernova remnants (SNRs) 
represent a class of SNRs which consist of thermal X-ray-filled centers 
and radio-bright shells, such as W44, \snr, 3C391, and W49B (Rho \& 
Petre 1998; Jones \etal\ 1998).
This category of SNRs displays a good correlation with \HI\ or 
molecular clouds (MCs; Rho \& Petre 1998; also see the SNR--MC 
association catalog in Jiang \etal\ 2010) and shows a strong 
association with 
1720 MHz OH masers (Green \etal\ 1997; Yusef-Zadeh \etal\ 2003).
The MM~SNRs have received two decades of studies since the standard 
Sedov (1959) evolution fails to predict their centrally peaked 
X-ray morphology. 
Debates on the origin of the central X-rays are still on-going.
Several competitive scenarios were proposed to explain the bright 
X-ray emission in the MM~SNR interior. 
The first one invokes the thermal conduction model leading to a central 
density increase (Cui \& Cox 1992; Shelton \etal\ 1999; Cox \etal\ 1999).
It predicts that the temperature smoothly decreases while the density 
rises away from the center, creating a ``center-filled" hot X-ray 
morphology with a thin, dense, and cold radiative shell in the periphery. 
The second scenario is the cloudlet evaporation model (White \& Long 1991).
In this model, ambient cloudlets engulfed by the SNR shock gradually 
evaporate into the hot gas, resulting in an increased density in the
SNR interior.

Recent \Suzaku\ observations of a handful MM~SNRs (IC443, W49B,
G359.1$-$0.5) have detected strong radiative recombination continua 
(RRCs), giving the first direct evidence of over-ionization in SNRs 
(Yamaguchi \etal\ 2009; Ozawa \etal\ 2009; Ohnishi \etal\ 2011 ).
In these SNRs, the ionization temperature $\kTz$ has been found
to be significantly higher than the electron temperature $kT$.
As for the origin of the recombining plasma, Itoh \& Masai (1989)
discussed the rapid electron cooling caused by the rarefaction process 
after the shock breaks out of the dense circumstellar medium to 
the rarefied ambient medium.
Kawasaki \etal\ (2002) proposed thermal conduction as a mechanism to
produce the over-ionized plasma.
Zhou \etal\ (2011) used a two-dimensional model to investigate 
SNRs evolving in a ring-like cloud and found that the thermal 
conduction and the rapid adiabatic expansion are two classes of 
processes both responsible for the over-ionized plasma in W49B.

In the \rray\ band, a new generation of telescopes such as the 
High-Energy Stereoscopic System (H.E.S.S) and {\it Fermi} detected 
many new 
SNR--\rray\ associations (Ferrand \& Safi-Harb 
2012\footnotemark[5]\footnotetext[5]{\url{http://www.physics.umanitoba.ca/snr/SNRcat}}
and references therein), providing the best database for studying cosmic rays 
acceleration in SNRs to the very high energy (VHE) band.
Although distinguishing the hadronic \rray\ emission from the leptonic 
emission is still an on-going effort, accumulating evidence has pointed 
to SNRs interacting with MCs emitting hadronic \rray s 
via proton--proton collision (such as W28, IC443, W51C; Abdo \etal\ 
2009, 2010; Li \& Chen 2012).
MM~SNRs are ideal sites to study hadronic cosmic rays in the VHE band
due to its good correlation with the 1720 MHz OH masers, as the masers 
are considered to be new potential tracers for the sites with hadronic 
particle acceleration (Hewitt \etal\ 2009; Frail 2011).
Furthermore, in terms of statistics, recent \Fermi-LAT observations have 
revealed that MM~SNRs occupy a considerable percentage of the \rray-bright 
SNRs within middle-age (Hewitt \& Fermi LAT Collaboration 2012) and 
most of those MM~SNRs likely/probably emit \rray s via the hadronic 
process (Brandt \& Fermi LAT Collaboration 2012).

Cosmic ray interactions with a dense medium can generate non-thermal 
radiation in a broad band.
Several models have been established to predict the broad band spectrum 
in MC--SNR interaction sites (e.g.\ Bykov \etal\ 2000; Gabici 
\etal\ 2009).
However, non-thermal X-ray studies of SNRs in the \rray-bright 
and MC--SNR interaction sites are still limited.
MM~SNRs, especially those bright in both X-ray and \rray\ bands, are good 
candidates to explore the non-thermal X-ray emission.

\snr\ (G6.4$-$0.1) is an evolved SNR, with an estimated age 3.3--$4.2
\E{4}$ yr (Rho \& Borkowski 2002, hereafter RB02; Vel{\'a}zquez 
\etal 2002; Li \& Chen 2010) at a distance of about 2 kpc 
(Ilovaisky \& Lequex 1972; Goudis 1976; Vel{\'a}zquez \etal\ 2002).
It is a well-known MM~SNR with double radio shells in its northern
hemisphere and a breakout morphology in its southwest (RB02).
The interaction of the MCs with the blast wave of the SNR has been 
studied for 30 yr.
Wootten (1981) firstly found broadenings of CO and HCO$^+$ lines in the 
northeastern (NE) MC and suggested the cloud is compressed and heated by 
the \snr\ shock.
The interaction was confirmed with the detections of 1720 OH masers 
in the NE MC and the cloud on the inner radio shell (Frail \etal\ 1994; 
Claussen \etal\ 1997).
Follow-up (sub)millimeter and infrared observations toward the SNR 
(Arikawa \etal\ 1999; Reach \etal\ 2005; Nicholas \etal\ 2012)
gave further evidence for the MC--SNR interaction, and provided detailed 
physical properties of the smaller clumps embedded in the MCs.

As a prominent \rray-emitting SNR, \snr\ was detected at energy
$>300$ MeV by COS-B (Pollock 1985), 
$>100$ MeV by EGRET (Sturner \& Dermer 1995; Hartman \etal\ 1999),
$>400$ MeV by {\it AGILE} (Giuliani \etal\ 2010),
$>0.2$ GeV by Fermi (Abdo \etal\ 2010),
and $>0.1$ TeV by H.E.S.S (Aharonian \etal\ 2008).
The GeV--TeV \rray s are considered to be produced by hadronic 
interaction of cosmic rays with the NE MCs
(Aharonian \etal\ 2008; Abdo \etal\ 2010; Li \& Chen 2010; 
Ohira \etal\ 2011; Yan \etal\ 2012).

In the X-ray band, \snr\ has been observed with several generations of 
X-ray satellites, however, inconsistent spectral results have been 
obtained.
Early \Einstein\ IPC $+$ MPC analysis has shown that \snr\ is a composite 
SNR and a single thermal component was used to characterize the X-ray 
emission (Long \etal\ 1991).
The \Chandra\ X-ray imaging revealed a bright knotty region in the SNR 
interior (Keohane \etal\ 2005).
Combining the observations of \ROSAT\ and \ASCA, RB02 showed that a 
two-thermal-component model (with temperatures of 0.6 and 1.8 keV) is 
required to reproduce the global X-ray spectrum from the SNR center.
This conclusion was also supported by a subsequent \ASCA\ study 
(Kawasaki \etal\ 2005; hereafter KO05), which searched for evidence 
of over-ionized plasma but without a successful detection.
However, a most recent \Suzaku\ analysis disfavored the presence of the
hot (1.8 keV) plasma and argued that the central X-rays are emitted 
by recombining plasma with an electron temperature $\sim 0.4$ keV
(Sawada \& Koyama 2012; hereafter SK12).
Hence, the nature of the central X-ray emission of \snr\ is still 
controversial.

Furthermore, for the NE shell where \rray\ emission has been detected,
there is debate on the nature of its X-ray emission, in particular
whether there is any non-thermal component or not.
RB02 applied a thermal component (0.56 keV) to characterize the X-ray 
spectrum observed with \ROSAT\ and \ASCA, while Ueno \etal\ 
(2003) described the X-ray emission by a two-component model consisting
of a cold ($kT\sim 0.3$ keV) thermal plasma and a very hard 
($\Gamma=1.3^{+0.6}_{-0.9}$) non-thermal tail ($apec$ + \powerlaw)
using an 18 ks EPIC pn observation with \XMMN.
Testing the existence of the non-thermal X-ray emission and investigating
its origin is essential, 
because X-rays are in the lower neighboring energy band of 
\rray s, and thus provide precious information for the broad-band study of 
cosmic ray acceleration in SNRs.

Motivated by the inconsistent X-ray results previously obtained with
different telescopes, the need to verify and study the non-thermal
emission and to search for any spatial variations of the physical 
properties across \snr, we here present our X-ray analysis 
based on \XMMN\ observations.  In Section~\ref{S:obs}, 
we describe the observations and the data calibration.  
Section~\ref{S:result} shows the X-ray image and the spatially 
resolved spectroscopic results.  
In Section~\ref{S:discussion}, we discuss the physical properties of 
the hot gas inside \snr\ with a comparison to previous X-ray studies, 
and the origin of the non-thermal X-ray emission detected in the NE 
shell. The conclusions are summarized in Section~\ref{S:summary}.

\section{Observations and data reduction} \label{S:obs}
Four \XMMN\ observations toward the SNR \snr\ were performed 
in three pointings and cover the northern hemisphere of the remnant.
Two observations of the NE shell were carried out on 2002 September 
23 and on 2003 October 7--8, respectively, in full frame mode and 
with thick filter (PI: K. Koyama).  
Another two archival observations pointing to the center and north of 
\snr\ were obtained from the observations toward the Galactic ridge, 
which were made on 2003 March 19--20 and 2003 March 20, respectively, 
in full frame mode and with medium filter (PI: R. Warwick).
Detailed information of the \XMMN\ data, including the coordinates 
and the effective exposure times (after removing time intervals with 
heavy proton flarings) of MOS (Turner \etal\ 2001) and pn (Str{\"u}der 
\etal\ 2001) for the NE shell, SNR center, and northern region are 
summarized in Table~\ref{T:obs}.
The Science Analysis System software (SAS,\footnotemark[6]
\footnotetext[6]
{\url {http://xmm.esac.esa.int/sas/}} ver. 11.0) was used to reprocess 
the EPIC data.

\begin{deluxetable*}{p{3cm}ccccccc}
\tabletypesize{\footnotesize}
\tablecaption{Summary of the \XMMN\ observations of the SNR \snr}
\tablewidth{0pt}
\tablehead{
\colhead{Pointing} & {Obs. ID}  & Obs. Date &\multicolumn{2}{c}{Center (J2000)}  & & \multicolumn{2}{c}{Exposure$^{\rm a}$ (ks)} \\ \cline{4-5} \cline{7-8}
&  & & R.A. & Decl. & & MOS & pn
}
\startdata
Center \dotfill  & 0135742201 & 2003 Mar 19--20 & 18:00:21.1 & $-$23:20:19.0 & & 5 & 4 \\
North  \dotfill  & 0135742501 & 2003 Mar 20 & 18:01:04.3 & $-$23:02:57.0 & & 1 & 1 \\ 
Northeast \dotfill  & 0145970101 & 2002 Sep 23 & 18:01:45.0 & $-$23:18:00.0 & & 21 & 15 \\
Northeast \dotfill  & 0145970401 & 2003 Oct 7--8 & 18:01:45.0 & $-$23:18:00.0 & & 9 & 9
\enddata
  \tablenotetext{a}{\phantom{0} Effective exposure time.}
\label{T:obs}
\end{deluxetable*}

\section{Results} \label{S:result}

\subsection{X-ray image}\label{S:image}
\begin{figure}[t]
\hspace{-0.07in}
\centerline{ {\hfil\hfil
\psfig{figure=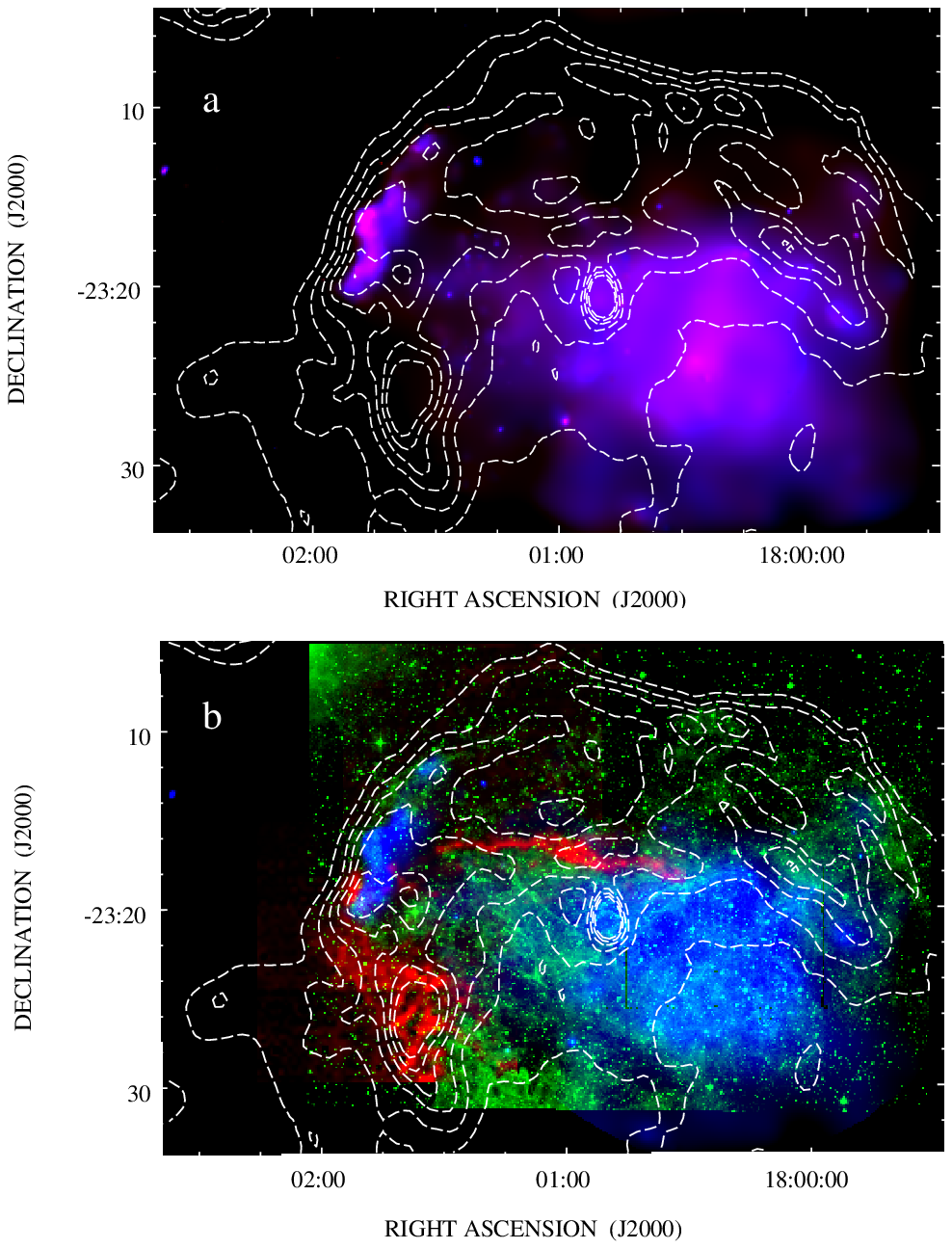,width=3.59in,angle=0, clip=}
\hfil\hfil}}
\caption{
(a) \XMMN\ EPIC color image of the SNR \snr, with red and blue 
colors corresponding to the 0.3--1.0 keV and 1.0--7.0 keV bands, respectively.
Radio contours are overlaid using VLA 1.4 GHz continuum emission 
with levels of 0.1, 0.22, 0.34, 0.46, 0.58, 0.7 Jy beam$^{-1}$.
(b)
Tri-color image of the SNR \snr. Red: the intensity map of JCMT 
\twCO~\Jttt\ integrated from $-$40 \kms\ to 40 \kms\ (Arikawa 
\etal\ 1999);
green: the narrow-band \Ha\ image obtained from the archival SuperCOSMOS 
\Ha\ Survey (Parker \etal\ 2005);
and blue: \XMMN\ 0.3--7.0 keV X-ray map. The 1.4 GHz radio contours are 
shown with the same levels as those in panel a.
}
\label{f:ximg}
\end{figure}

From the available \XMMN\ data, we generated a two-color (0.3--1.0 keV 
in red and 1.0--7.0 keV in blue) energy map covering the northern 
hemisphere of \snr, overlaid with 1.4 GHz radio contours (as shown 
in Figure~\ref{f:ximg}(a)).
Each of the images in the two bands was exposure-corrected and 
adaptively smoothed (using ``csmooth" command in CIAO\footnotemark[7]
\footnotetext[7]
{\url{http://cxc.harvard.edu/ciao/}} 4.3) with a Gaussian kernel 
$\sigma$ to achieve a minimum and maximum significance of 3 and 4, 
respectively.

The brightest X-ray emission arises from the NE shell and the central 
region as delineated by the inner radio shell.
Fainter and diffuse X-ray emission is also seen in the gap between 
the two radio shells.
A section of the radio shell in the northwestern edge is partially 
X-ray brightened. 
On the NE shell, an X-ray slab appears distorted and in general 
bearing a morphological resemblance to the radio brightness peak.
The hot gas in the SNR interior appears blobby rather than homogeneous, 
which is also consistent with the clumpy nature suggested by the 
\Chandra\ imaging study (Keohane \etal\ 2005).

\subsection{The environmental gas}
We produce a tri-color image of the northern hemisphere of
\snr, illustrating the \twCO~\Jttt\ (red), \Ha\ (green), and
the X-ray (blue) morphology of \snr, overlaid with
the radio contours, for comparison (see Figure~\ref{f:ximg}(b). 
The figure shows that \snr\ is evolving in a complicated environment 
with a dense MC in the east and a molecular stripe on the northern 
inner shell.
There is a range of temperature and density distribution on 
the shells, represented by three kinds of emission:
cold and dense MCs indicated by CO emission located in the outermost 
boundary; hot and tenuous X-rays generally located behind; 
warm and moderately dense gas radiating \Ha\ emission and situated 
between the MCs and the X-ray-emitting gas.

The SNR interior filled by X-rays is also bright in diffuse, nonuniform 
\Ha\ emission, whereas filamentary \Ha\ structures are mostly aligned 
with the inner radio shell or distributed in the east.
The \Ha\ filaments inside the SNR, in addition to the boundary of the 
diffuse \Ha\ emission in the south and west, appear to confine the 
central bright X-rays.

\begin{figure}[t]
\centerline{ {\hfil\hfil
\psfig{figure=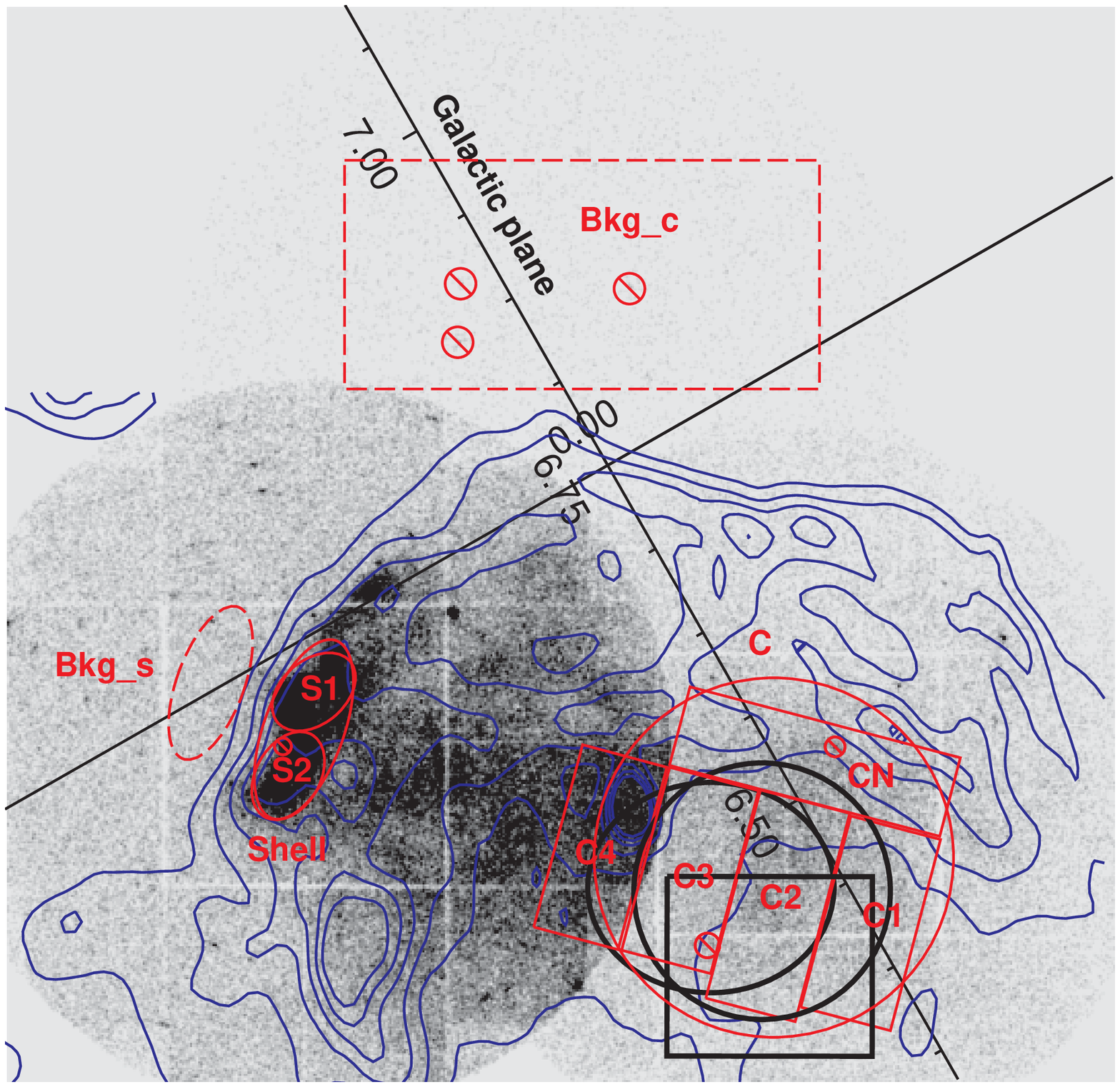,height=3.45in,angle=0, clip=}
\hfil\hfil}}
\caption{
\XMMN\ raw image of \snr\ overlaid with radio contours (at the same 
levels as those in Figure~\ref{f:ximg}(a)).
The thin solid and dashed regions colored in red correspond to source 
and background spectral extraction regions of \XMMN\ data.
The background spectra for the NE and central observations are 
extracted separately from the dashed ellipse and northern rectangle.
The thin solid circle (``C'') in red covers an area that coincides 
with that chosen by RB02 for spectral analysis of the central gas.
The thick black rectangle and ellipse indicate the spectral extraction 
regions of the ``center'' and ``east'' region, respectively, in KO05.
The spectral extraction region in SK12 is shown with the thick black 
circle. The location of the Galactic plane is also shown.
}
\label{f:reg}
\end{figure}

\subsection{\XMMN\ Spectral Analysis}

Prior to the extraction of spectra, we removed the point-like sources 
detected in the extraction regions from the events files.
For the three observations pointing to the northeast and center of the 
remnant, we applied a combination of source detection algorithms 
(wavelet, sliding-box, and maximum likelihood centroid fitting) 
described in Wang \etal\ (2003).
For the Galactic ridge observation pointing to the north of the remnant
(Obs. ID: 0135742501) which with a short exposure ($\sim 1$ ks), we 
detected point-like sources with the SAS edetect\_chain script and  
used a subtraction radius of $0.6'$.

\begin{figure}[t]
\vspace{0.1in}
\hspace{-0.13in}
\centerline{ {\hfil\hfil
\psfig{figure=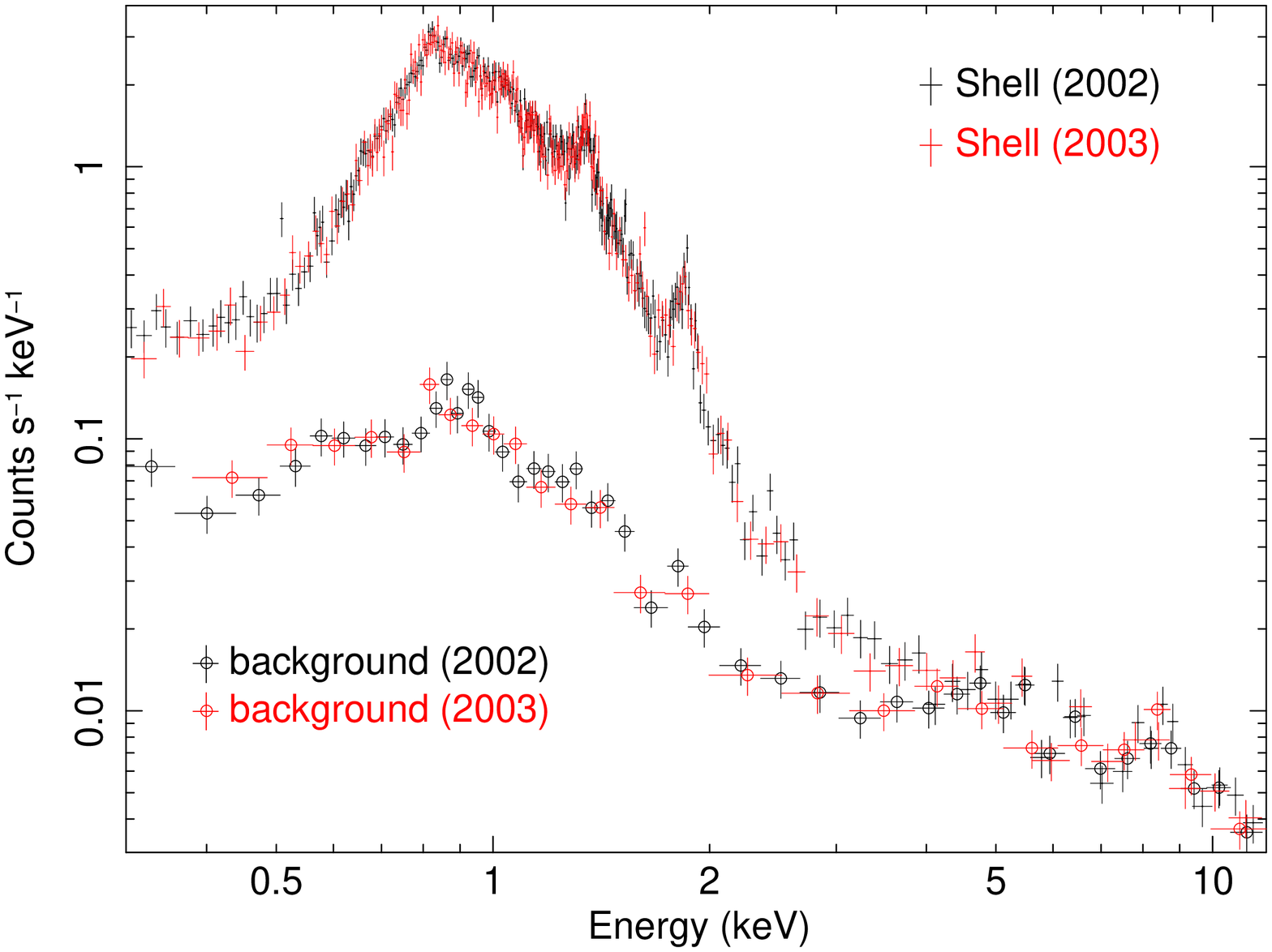,width=3.53in,angle=0, clip=}
\hfil\hfil}}
\caption{
EPIC-pn spectra of region ``Shell'' without background subtraction
(dots with error bars) and the normalized background (circles with error bars). 
We colored the 2002 (Obs ID: 0145970101) and 2003 (Obs ID: 0145970401) 
data in black and red, respectively.
Each spectrum is grouped with a minimum of 40 counts per bin.
The background-subtracted spectra are shown in Figure~\ref{f:sspec}.
}
\label{f:shell_bkg}
\end{figure}

We defined two large regions (``Shell'' and ``C'') covering the 
NE shell and central gas, respectively, to investigate the mean physical 
characteristics of the two regions with high statistics, and seven
small regions (``S1'', ``S2'', ``CN'', ``C1'', ``C2'', ``C3'', 
``C4''\footnotemark[8]
\footnotetext[8]{The region ``C4'' is in the field of the observation pointing
to the SNR center and is also covered by MOS1 of the two NE observations.}
) along the outer shell and inside the SNR to search 
for spectral variation (see Figure~\ref{f:reg}).
The extraction of the background spectra is described in 
Section~\ref{S:bkg}.
Each individual spectrum is adaptively binned to achieve a 
background-subtracted signal-to-noise ratio (S/N) of 3,
except that the spectra shown in Figure~\ref{f:shell_bkg} are
grouped to achieve a minimum of 40 counts per bin.

The \vmekal, \vnei, and \powerlaw\ models in the
XSPEC\footnotemark[9]\footnotetext[9]{\url{http://xspec.gsfc.nasa.gov}}
package (ver. 12.7) and the \neij\ model in the SPEX software 
(ver. 2.03.03; Kaastra \etal\ 1996) were used for spectral fitting
(see Sections~\ref{S:NEresult} and \ref{S:result_cent} for 
detailed description of the models).
The spectra of EPIC MOS1, MOS2 and pn from all available observations are 
jointly fitted together for the individual regions to better constrain 
the physical parameters.

\subsubsection{Background Subtraction} \label{S:bkg}
\snr\ is an extended object ($\sim 50'$ in diameter) located in the 
Galactic plane and close to the Galactic center, where the X-ray emission
from the SNR suffers contamination from a high Galactic X-ray 
background (SK12).
Different background regions were selected in the previous \ASCA\ and 
\Suzaku\ studies and inconsistent results were 
obtained (see Figure~\ref{f:reg} for the spectral extraction regions
used in this study and by RB02, KO05 and SK12). 
In Section~\ref{S:comp} below we provide a detailed comparison between 
the earlier studies and our study.
An appropriate background selection is therefore crucial, especially for 
a reliable study of the central gas that is more than $14'$ away 
from the northern boundary.

Accordingly, we defined two separate background regions, ``Bkg\_s" and 
``Bkg\_c'' (see Figure~\ref{f:reg}), for each of the NE observations
and the central observation of \snr, respectively.
Region ``Bkg\_s'' was selected as close as possible to 
the NE shell and outside the remnant's radio boundary.\footnotemark[10]\footnotetext[10]
{Region ``Bkg\_s'' was selected on the same MOS's central CCDs that 
also cover the source region ``Shell'' to ensure ``Bkg\_s'' has 
similar sky and instrumental background to that of the shell regions.
The vignetting effect is not expected to be important for the
shell considering that ``Bkg\_s'' is adjacent ($4\arcmin$) to
the ``Shell''.}
As the field of view of the \XMMN\ observation toward the SNR center 
(Obs. ID: 0135742201) is almost filled by the remnant, we do not find 
a proper background region from the same observation.
Therefore, ``Bkg\_c'' is selected from a nearby observation 
(Obs.\ ID: 0135742501), which pointed to the {\it same} Galactic 
latitude as the central gas and was fortunately performed with the same 
telescope configuration one day afterward.
Region ``Bkg\_c" is the closest background to the remnant and the most 
appropriate one to minimize any contamination from the Galactic ridge.

\subsubsection{Northeastern Shell}\label{S:NEresult}

\snr\ is a distinct MM~SNR partly due to its bright, deformed X-ray 
shell along the NE boundary.
Figure~\ref{f:shell_bkg} shows the pn spectra extracted from the region 
``Shell'' and its nearby background (normalized by area).
The X-ray spectra of the shell show several line features, such as
Mg He$\alpha$ ($\sim 1.35$~keV), Si He$\alpha$ ($\sim 1.86$~keV), 
and S He$\alpha$ ($\sim 2.45$~keV), supporting a thermal
plasma origin for at least part of the X-ray emission.
The shell X-ray emission extends up to 5 keV, while the background 
dominates the emission above 5~keV.

\begin{figure}[t]
\vspace{0.07in}
\hspace{-0.12in}
\centerline{ {\hfil\hfil
\psfig{figure=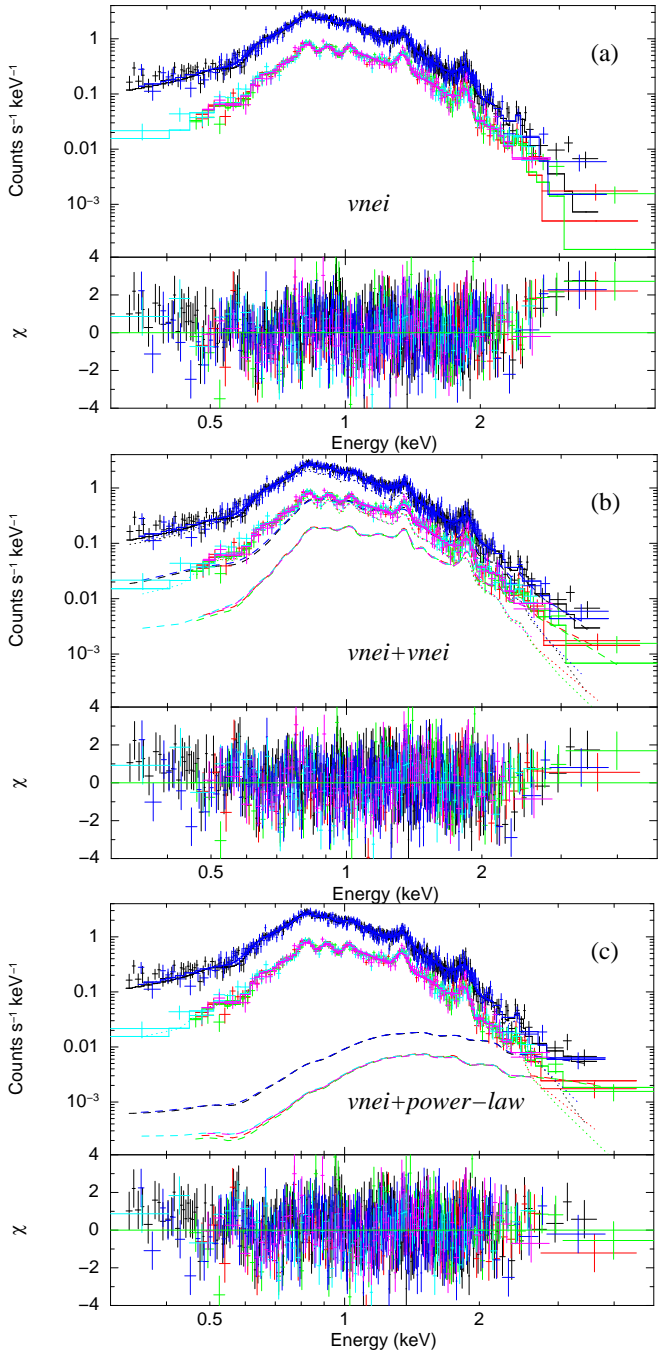,width=3.58in,angle=0, clip=}
\hfil\hfil}}
\caption{
\XMMN\ EPIC spectra of the NE shell (region ``Shell'') fitted with absorbed 
\vnei\ (a), \vnei+\vnei\ (b), and \vnei+\powerlaw\ (c) models.
Each individual spectrum is adaptively binned to achieve a 
background-subtracted S/N of 3.
Upper solid lines show pn spectra in black (observed in 2002) and blue 
(observed in 2003). 
Lower solid lines show MOS spectra colored in red, green (both 
observed in 2002), cyan and pink (both observed
in 2003).
The short and long dashed lines show the soft and hard components of the
model, respectively.
}
\label{f:sspec}
\end{figure}

Figure~\ref{f:sspec} shows the EPIC (MOS+pn) background-subtracted 
spectra of the shell.
Adopting the cross-sections from Morrison \& McCammon (1983) for the 
foreground absorption, an absorbed single collisional ionization 
equilibrium (CIE) model, \vmekal, and an absorbed non-equilibrium 
ionization (NEI) model, \vnei\ in XSPEC were initially applied to 
interpret the emission in the large NE region ``Shell". 
We allow the abundances of C, Si, S, and Fe to vary, and tie the 
abundances of all other elements (N, O, Ne, Mg, Ar, Ca, and Ni) to 
that of C.
The absorbed \vmekal\ model does not produce a good fit
to the soft X-ray spectra ($\chi_{\nu}^2\sim 1.75$), while the \vnei\ 
model with a plasma temperature $kT=0.33\pm0.1$~keV appears to provide a 
better fit ($\chi_{\nu}^2\sim 1.33$); however
it seriously underestimates the flux above 2.5~keV as shown in 
Figure~\ref{f:sspec}(a).
We therefore added another component to account for the hard X-ray 
tail in the spectra; and the addition can be either a second NEI plasma
model or a \powerlaw\ one.
The obtained $\chi_{\nu}^2$ values are $1.22$ and $1.28$, 
respectively, corresponding to {\it F}-test null hypothesis probabilities
of  $1.1\E{-17}$ and $3.6\E{-10}$ for adding the hard components,
which supports the need for a two-component model.

\setlength{\tabcolsep}{2pt}
\begin{deluxetable*}{p{3.5cm}cccccccccccc}
\tabletypesize{\scriptsize}
\tablecaption{Spectral fitting results for the northeastern shell 
of \snr\ with 90\%\ confidence}
\tablewidth{0pt}
\tablehead{
\colhead{Regions} & \multicolumn{3}{c}{Shell} & & \multicolumn{2}{c}{S1} 
& & \multicolumn{2}{c}{S2}  \\ \cline{2-4} \cline{6-7} \cline{9-10}
\colhead{Models} & \vnei & \vnei+\vnei$^{\rm a}$ & \vnei+\powerlaw & 
 & \vnei+\vnei$^{\rm a}$ & \vnei+\powerlaw &
 & \vnei+\vnei$^{\rm a}$ & \vnei+\powerlaw
}
\startdata
$\chi_{\nu}^{2}$ (dof) \dotfill
 & 1.33 (1019)
 & 1.23 (1016) & 1.28 (1017) &
 & 1.12 (833) & 1.14 (834) &
 & 1.17 (783) & 1.19 (784) \\
$\NH$ ($10^{21}\cm^{-2}$) \dotfill  
 & $7.87^{+0.23}_{-0.27}$
 & $7.76^{+0.31}_{-0.13}$ & $8.15^{+0.25}_{-0.23}$ &
 & $7.93^{+0.37}_{-0.34}$ & $8.19^{+0.33}_{-0.32}$ &
 & $7.50^{+0.48}_{-0.44}$ & $7.63^{+0.35}_{-0.35}$ \\
\hline
\colhead{Soft component} \\
$\kTc$ ($\keV$) \dotfill
 & $0.33^{+0.01}_{-0.01}$
 & $0.29^{+0.01}_{-0.01}$ & $0.30^{+0.01}_{-0.01}$ &
 & $0.29^{+0.01}_{-0.01}$ & $0.30^{+0.02}_{-0.01}$  &
 & $0.29^{+0.02}_{-0.01}$ & $0.31^{+0.02}_{-0.01}$ \\
$\tau_{\rm c}$ ($10^{12}\cm^{-3}\,{\rm s}$) \dotfill
 & $0.60^{+0.15}_{-0.20}$
 & $<500$ & 1.06 ($>0.76$) &
 & $<500$ & $1.14 (> 0.56)$ &
 & $<500$ & $1.10 (> 0.74)$\\
Abun$^{\rm b}$ \dotfill
 & $0.26^{+0.03}_{-0.03}$
 & $0.44^{+0.10}_{-0.09}$ & $0.33^{+0.11}_{-0.06}$ &
 & $0.47^{+0.17}_{-0.11}$ & $0.34^{+0.16}_{-0.07}$ &
 & $0.34^{+0.10}_{-0.09}$ & $0.27^{+0.08}_{-0.05}$ \\
${\rm [Si/H]}$ \dotfill
 & $0.40^{+0.08}_{-0.07}$
 & $0.43^{+0.08}_{-0.07}$ & $0.59^{+0.18}_{-0.10}$ &
 & $0.50^{+0.11}_{-0.08}$ & $0.61^{+0.07}_{-0.14}$ &
 & $0.34^{+0.09}_{-0.09}$ & $0.43^{+0.14}_{-0.10}$\\
${\rm [S/H]}$ \dotfill
 & $0.88^{+0.30}_{-0.27}$
 & $0.37^{+0.20}_{-0.17}$ & $1.04^{+0.30}_{-0.38}$ &
 & $0.46^{+0.28}_{-0.24}$ & $1.11^{+0.43}_{-0.46}$ &
 & $0.44^{+0.27}_{-0.24}$ & $1.04^{+0.42}_{-0.43}$ \\
${\rm [Fe/H]}$ \dotfill
 & $0.22^{+0.03}_{-0.03}$
 & $0.41^{+0.07}_{-0.07}$ & $0.29^{+0.02}_{-0.05}$  &
 & $0.45^{+0.15}_{-0.11}$ & $0.30^{+0.17}_{-0.07}$ &
 & $0.34^{+0.08}_{-0.08}$ & $0.23^{+0.05}_{-0.06}$ \\
$F_{\rm s}$~($10^{-11}\erg\cm^{-2}\ps$)$^{\rm c}$  \dotfill
 & 6.61 & 4.98 & 7.72  &
 & 2.72 & 3.86 &
 & 1.89 & 2.52 \\
\hline
\colhead{Hard component} \\
$\kTh$ ($\keV$) \dotfill
 & $\cdots$
 & $0.61^{+0.07}_{-0.06}$ & $\cdots$ &
 & $0.63^{+0.06}_{-0.06}$ & $\cdots$ &
 & $0.65^{+0.34}_{-0.08}$& $\cdots$ \\
$\tau_{\rm h}$ ($10^{12}\cm^{-3}\,{\rm s}$) \dotfill
 & $\cdots$
 & $0.79^{+0.85}_{-0.33}$ & $\cdots$ &
 & 0.80 ($> 0.35$) & $\cdots$ &
 & $0.44^{+0.55}_{-0.32}$ & $\cdots$ \\
$\Gamma$ \dotfill
 & $\cdots$
 & $\cdots$ & $1.85^{+0.54}_{-0.98}$ &
 & $\cdots$ & $0.24^{+1.87}_{-2.25}$ &
 & $\cdots$ & $1.32^{+1.04}_{-1.81}$\\
$F_{\rm h}$~($10^{-12}\erg\cm^{-2}\ps$)$^{\rm c}$  \dotfill
 & $\cdots$ & 7.59 & 0.34 &
 & 3.01 & 0.11  &
 & 2.20 & 0.09 
\enddata
\tablecomments{
Three models are applied to fit the spectra of the NE region
``Shell''. Two-component models better reproduce the hard X-ray emission and
are thus subsequently used for the small-scale regions ``S1'' and ``S2''.
The parameters for the hard components show the fit results of the two-component
models, where the temperature $\kTh$ and ionization timescale $\tau_{\rm h}$ 
are only applied for the hot component in the \vnei+\vnei\ model, 
while $\Gamma$ is the photon index of \powerlaw\ component.
See the spectra of region ``Shell'' in Figure~\ref{f:sspec}.
}
  \tablenotetext{a}{\phantom{0}
The abundances of the soft and hard component are coupled.
}
  \tablenotetext{b}{\phantom{0} The tied abundances of C, N, O, Ne, Mg,
Ar, Ca, and Ni. }
  \tablenotetext{c}{\phantom{0} The best-fitted unabsorbed fluxes 
$F_{\rm s}$ (soft component) and $F_{\rm h}$ (hard component) in 
the 0.3--5.0 keV range.}
\label{T:shell}
\end{deluxetable*}

The spectra fitted with the absorbed single \vnei, double \vnei, 
and \vnei +\powerlaw\ models are shown in Figure~\ref{f:sspec} and 
the fitting results are tabulated in Table~\ref{T:shell}.
The three models commonly support the origin of the soft X-ray 
emission being a $\sim 0.3$~keV plasma with under-solar abundances
(except for [S/H]$\sim1$ in the \vnei+\powerlaw\ model).
The double-\vnei\ model reproduces the spectra well in the soft X-ray band 
($\le3$~keV) and gives the best fit among the three models in view of 
$\chi_\nu^2$. 
The cold component has a temperature $\kTc\sim 0.3$~keV (with ionization 
timescale $\tau_c$ poorly constrained), while the hot component has a 
temperature $\kTh\sim 0.6$~keV and is under-ionized ($\tau_h\sim 
1\E{12} \cm^{-3} \s$).
The \vnei+\powerlaw\ model also reproduces well the X-ray spectra of the 
NE shell. 
It is actually reasonable to search in the shell for non-thermal X-rays that 
are produced by the electrons accelerated at the shock sites, considering 
that the shell is spatially correlated 
with the \rray-emitting area that interacts with the MC.
In this model, the X-ray emission from the shell consists of a thermal 
component characterized by a relatively cold ($kT=0.3$~keV) plasma 
with ionization timescale $\tau_c>7.6\E{11}~\cm^{-3}$~s and low metal 
abundances (except for S), plus a non-thermal component with a hard 
photon index ($\Gamma=1.9^{+0.5}_{-1.0}$).
Compared to the double-\vnei\ model, the \vnei+\powerlaw\ model fits 
the hard X-ray band (3--5~keV) better\footnotemark[11]\footnotetext[11]{
We did not detect any source emission with S/N$\ge 3$ between 5 and 8 keV.
The best-fit results of the \vnei+\powerlaw\ model gives a flux of 
$\sim 7\E{-14}~\erg\cm^{-2}\ps$ in the 5--8~keV range, which is, however,
about an order of magnitude smaller than the background level.
This explains the lack of background-subtracted spectral bins above
5~keV.
We also simulated the spectra using the best-fit results of the \vnei+
\powerlaw\ model to check if we can obtain a few bins above 5~keV with 
the given response files and exposure time.
Only two bins are obtained in the MOS spectra but they span from 
$<4$~keV to 8~keV thus are not statistically important.
}
but has slightly larger residuals in the soft band.
Note that there are only a few bins above 3 keV, which play a minor
role in affecting the $\chi_\nu^2$ value. 
The lack of sufficient counts above 3 keV does not allow us to 
distinguish between the two models. Deeper observations are needed, 
especially to improve on the statistics of the hard X-ray emission 
above $\sim 3$~keV.

We further examine the spectral properties of two separate regions,``S1" 
and ``S2", along the shell, in light of the deformed morphology of 
the shell and the non-uniform environment at the NE boundary.
We also applied the double-\vnei\ and \vnei+\powerlaw\ models to fit the 
spectra and the fitting results are summarized in Table~\ref{T:shell}. 
The X-ray emission of the small-scale regions share similar spectral
properties to those of the larger region ``Shell'' based on the 
available \XMMN\ data.

\subsubsection{Central Gas} \label{S:result_cent}
We first focus on the larger central region ``C" which encircles the 
brightest X-ray-emitting gas in the SNR interior. 
This region has a similar sky coverage with the ``center" region defined 
in RB02.
There are three distinct line features, Mg \Hea, Si \Hea, and S \Hea,
and several weaker bumps contributed from K$\alpha$ lines such as H-like 
Mg ($\sim 1.47$ keV) and Si ($\sim 2.0$ keV), confirming the thermal 
origin of the X-ray emission.
We ignore the bins above 5 keV in the following spectral fitting of 
the central X-ray because of the poor statistics and the spectrum being
background dominated.
Although an excess is seen above 5 keV in both pn and MOS spectra,
we do not discuss its nature in this paper given that the strength of 
this excess is not statistically significant.

Single-temperature CIE (\vmekal) and NEI (\vnei) models were first
applied to the central region.
We allow the abundances of C, Mg, Si, S and Fe to vary, and tied 
the abundances of all other elements to C.
The fit results are summarized in the second and third columns of 
Table~\ref{T:center_s}.
The \vmekal\ and \vnei\ models give a similar foreground hydrogen 
column density ($\NH\sim4\E{21}~\cm^{-2}$) and plasma temperature 
($kT\sim0.6$~keV), which also agreeably support low elemental 
abundances of the hot plasma. 
However, we find that none of the \vmekal\ and \vnei\ models yield 
adequately good fits to the spectra ($\chi_{\nu}^2 \sim 
1.6$ and $\sim 1.7$, respectively) since they failed 
to account for the emission above 2.5 keV and for the weak bump at 
around 2.0~keV probably due to the \Lya\ line of Si.

To better reproduce the spectra, a two-component model \vnei+\vnei\ 
was subsequently fitted, with the abundances of the cold and hot 
components tied together. 
The two-temperature NEI model improved the fitting to some extent 
($\chi_{\nu}^{2} \sim 1.4$) and gave a low ionization timescale 
($\tauc \sim 2\E{11}~\cm^{-3}$~s) for the cold component. 
Whereas, a much higher ionization timescale ($\tauh \sim 
10^{12}~\cm^{-3}$~s with the upper limit being unconstrained) is 
obtained for the hot component, implying the hotter gas is (or 
close to) in ionization equilibrium.

\begin{figure*}[tbh!]
\hspace{-0.1in}
\centerline{ {\hfil\hfil
\psfig{figure=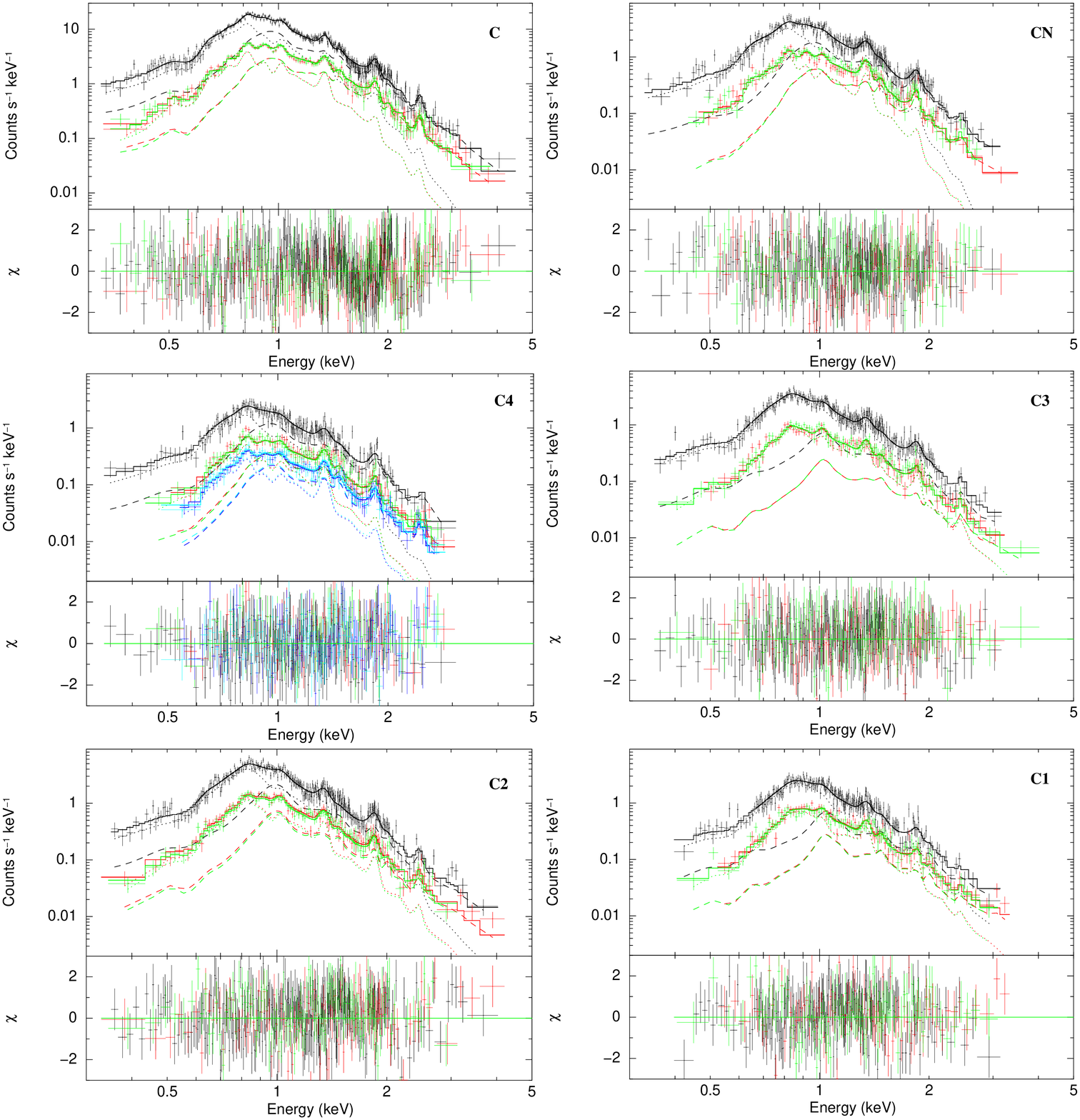,width=1.02\textwidth,angle=0, clip=}
\hfil\hfil}}
\caption{
\XMMN\ EPIC spectra and fitted model (absorbed \vnei+\vmekal) from the
six regions (indicated in Figure~\ref{f:reg}) in the SNR interior. 
Each individual spectrum is adaptively binned to achieve a 
background-subtracted S/N of 3.
The black solid lines show pn spectra, while MOS spectra are shown by 
the lower solid lines in red and green. The blue and cyan lines 
for region ``C4'' represent MOS1 spectra extracted from the southwestern 
edge of the NE observations with IDs 0145970101 and 0145970401, 
respectively.  
The short and long dashed lines show the soft and hard components of the
model, respectively.
}
\label{f:cspec}
\end{figure*}

We accordingly use a \vmekal\ model instead of \vnei\ to describe the 
hot component.
The \vnei+\vmekal\ model gives a satisfactory fit ($\chi_{\nu}^{2}
\sim 1.3$, see Table~\ref{T:center_d} and the upper-left panel of 
Figure~\ref{f:cspec}).
The spectral fit results show that, on average, the X-ray-emitting gas 
in the SNR interior has a low metal abundance (below solar values) and 
is composed of a cold ($\kTc\sim0.4$~keV) and under-ionized 
($\tauc \sim 4\E{11}~\cm^{-3}~\s$) plasma and a hot ($\kTh\sim0.8$~keV) 
plasma in CIE.

Given that the gas in the SNR interior is not uniform in either X-rays
or the other bands (see Figure~\ref{f:ximg}(b)) and region ``C" is large 
in scale (14$'$), we subsequently extracted five smaller regions 
(``CN", ``C1", ``C2", ``C3" and ``C4") to inspect any spatial variation 
of the gas properties in the central portion.
In Table~\ref{T:center_d} we summarize the \vnei+\vmekal\ fit results
of the five small-scale regions, and in Figure~\ref{f:cspec} we show 
the EPIC spectra with the fitted models.
As shown in the table, the absorption column density is generally  
increased from west to east, while the temperature is generally decreased. 
The highest absorption ($\NH\sim7\E{21}~\cm^{-2}$) and lowest 
temperature ($\kTc\sim 0.3$~keV) are found in the northern 
region ``CN'', which is on the inner shell and bright in CO and \Ha\ 
emission (see Figure~\ref{f:ximg}(b)).

\begin{deluxetable*}{lcccccccc}
\tabletypesize{\scriptsize}
\tablecaption{Spectral fitting results for the central gas of \snr\
with single-component models}
\tablewidth{0pt}
\tablehead{
\colhead{Models} & \vmekal & \vnei
 & \multicolumn{6}{c}{\neij} \\ \cline{4-9}
\colhead{Regions} & \colhead{C} & \colhead{C} & \colhead{C} 
& \colhead{CN} & \colhead{C4}
 & \colhead{C3} & \colhead{C2} & \colhead{C1} 
}
\startdata
$\chi_{\nu}^{2}$ (d.o.f.) \dotfill
 & 1.59 (622) & 1.69 (621)
 & 1.44 (620) & 1.44 (454) & 1.07 (539) 
 & 1.35 (438) & 1.18 (496) & 1.17 (392)\\ 
$\NH$ ($10^{21}\cm^{-2}$) \dotfill  
 & $4.05^{+0.09}_{-0.09}$ & $4.37^{+0.17}_{-0.10}$ 
 & $3.98^{+0.10}_{-0.09}$ & $3.84^{+0.21}_{-0.06}$ & $3.87^{+0.24}_{-0.08}$
 & $3.69^{+0.22}_{-0.20}$ & $4.11^{+0.20}_{-0.18}$ & $4.25^{+0.32}_{-0.09}$\\
$kT$ ($\keV$) \dotfill
 & $0.60^{+0.01}_{-0.01}$ & $0.60^{+0.01}_{-0.01}$
 & $0.59^{+0.01}_{-0.01}$ & $0.59^{+0.01}_{-0.02}$ & $0.57^{+0.02}_{-0.02}$
 & $0.54^{+0.01}_{-0.03}$ & $0.59^{+0.02}_{-0.01}$ & $0.63^{+0.04}_{-0.02}$\\
$\tau^{\rm a}$ ($10^{11}\cm^{-3}\,{\rm s}$) \dotfill
 & $\cdots$ & $2.35^{+0.21}_{-0.24}$ 
 & $8.58^{+0.61}_{-0.86}$ & $13.54^{+2.50}_{-3.19}$ & $8.20^{+0.40}_{-1.34}$
 & $8.66^{+0.63}_{-2.43}$ & $8.36^{+0.41}_{-1.22}$ & $7.45^{+3.59}_{-1.53}$\\
Abun$^{\rm b}$ \dotfill
 & $0.18^{+0.02}_{-0.02}$ & $0.12^{+0.01}_{-0.01}$
 & $0.22^{+0.02}_{-0.02}$ & $0.31^{+0.02}_{-0.06}$ & $0.22^{+0.03}_{-0.05}$
 & $0.13^{+0.02}_{-0.03}$ & $0.33^{+0.08}_{-0.06}$ & $0.29^{+0.03}_{-0.11}$\\
${\rm [Mg/H]}$ \dotfill
 & $0.26^{+0.03}_{-0.02}$ & $0.12^{+0.01}_{-0.01}$
 & $0.37^{+0.02}_{-0.04}$ & $0.60^{+0.05}_{-0.12}$ & $0.53^{+0.13}_{-0.09}$
 & $0.28^{+0.08}_{-0.04}$ & $0.46^{+0.05}_{-0.08}$ & $0.44^{+0.05}_{-0.10}$\\
${\rm [Si/H]}$ \dotfill
 & $0.18^{+0.02}_{-0.02}$ & $0.16^{+0.02}_{-0.02}$
 & $0.25^{+0.03}_{-0.02}$ & $0.29^{+0.04}_{-0.05}$ & $0.33^{+0.04}_{-0.05}$
 & $0.25^{+0.04}_{-0.04}$ & $0.34^{+0.04}_{-0.06}$ & $0.25^{+0.10}_{-0.05}$\\
${\rm [S/H]}$ \dotfill
 & $0.31^{+0.06}_{-0.06}$ & $0.39^{+0.07}_{-0.07}$
 & $0.32^{+0.07}_{-0.07}$ & $0.32^{+0.15}_{-0.15}$ & $0.51^{+0.15}_{-0.15}$
 & $0.47^{+0.13}_{-0.13}$ & $0.52^{+0.15}_{-0.14}$ & $0.36^{+0.19}_{-0.18}$\\
${\rm [Fe/H]}$ \dotfill
 & $0.12^{+0.01}_{-0.01}$ & $0.10^{+0.01}_{-0.01}$
 & $0.18^{+0.01}_{-0.01}$ & $0.21^{+0.02}_{-0.02}$ & $0.21^{+0.01}_{-0.04}$
 & $0.14^{+0.01}_{-0.02}$ & $0.26^{+0.03}_{-0.04}$ & $0.26^{+0.04}_{-0.05}$\\
$f\nel\nH V\du^{-2}~(10^{57}\cm^{-3})$ \dotfill  
 & $6.29^{+0.32}_{-0.31}$ & $7.45^{+0.35}_{-0.46}$
 & $5.91^{+0.38}_{-0.36}$ & $0.92^{+0.14}_{-0.12}$ & $0.78^{+0.13}_{-0.11}$
 & $1.21^{+0.23}_{-0.17}$ & $1.20^{+0.17}_{-0.15}$ & $0.88^{+0.16}_{-0.13}$\\
$\nH f^{1/2} \du^{1/2} (\cm^{-3})^{\rm c}$ \dotfill
 & $0.79^{+0.02}_{-0.02}$  & $0.86^{+0.02}_{-0.03}$
 & $0.77^{+0.02}_{-0.02}$ & $0.67^{+0.05}_{-0.04}$ & $0.88^{+0.07}_{-0.06}$
 & $1.07^{+0.10}_{-0.07}$ & $0.93^{+0.07}_{-0.06}$ & $0.87^{+0.08}_{-0.06}$ \\
$F$~($10^{-11} \erg\cm^{-2}\ps$)$^{\rm d}$  \dotfill
 & 8.99 & 10.43 & 8.72 & 1.63 & 1.23 & 1.49 & 2.17 & 1.50
\\
\hline
$kT_i$ ($\keV$)$^{\rm e}$ \dotfill
 & $\cdots$ & $\cdots$
 & $>6.49$ & $>1.34$ & $>2.20$
 & $>1.91$ & $>2.68$ & $>5.17$\\
 Average charge\\ 
~~~~Mg \dotfill 
 & $\cdots$ & $\cdots$ 
 & $11.11^{+0.06}_{-0.06}$ & $10.90^{+0.14}_{-0.12}$
 & $11.09^{+0.12}_{-0.09}$ & $11.01^{+0.19}_{-0.14}$
 & $11.10^{+0.11}_{-0.05}$ & $11.23^{+0.16}_{-0.20}$
\\
~~~~Si \dotfill 
 & $\cdots$ & $\cdots$ 
 & $12.65^{+0.09}_{-0.07}$ & $12.35^{+0.18}_{-0.16}$
 & $12.66^{+0.20}_{-0.14}$ & $12.58^{+0.26}_{-0.18}$
 & $12.66^{+0.14}_{-0.09}$ & $12.80^{+0.21}_{-0.31}$
\\
~~~~S \dotfill 
 & $\cdots$ & $\cdots$ 
 & $14.34^{+0.08}_{-0.06}$ & $14.09^{+0.14}_{-0.12}$
 & $14.35^{+0.16}_{-0.18}$ & $14.28^{+0.26}_{-0.21}$
 & $14.35^{+0.14}_{-0.13}$ & $14.48^{+0.22}_{-0.09}$
\\
~~~~Fe \dotfill 
 & $\cdots$ & $\cdots$ 
 & $18.08^{+0.36}_{-0.44}$ & $17.39^{+0.37}_{-0.20}$
 & $17.98^{+0.81}_{-0.83}$ & $17.55^{+1.17}_{-0.80}$
 & $18.07^{+0.68}_{-0.67}$ & $18.85^{+1.27}_{-1.23}$
\\
$\kTz$ (keV) \\ 
~~~~Mg \dotfill 
 & $\cdots$ & $\cdots$ 
 & $0.72^{+0.02}_{-0.02}$ & $0.65^{+0.05}_{-0.04}$
 & $0.71^{+0.03}_{-0.03}$ & $0.69^{+0.07}_{-0.05}$
 & $0.72^{+0.04}_{-0.02}$ & $0.77^{+0.08}_{-0.08}$
\\
~~~~Si \dotfill 
 & $\cdots$ & $\cdots$ 
 & $0.91^{+0.05}_{-0.03}$ & $0.75^{+0.10}_{-0.10}$
 & $0.92^{+0.10}_{-0.07}$ & $0.88^{+0.13}_{-0.10}$
 & $0.92^{+0.07}_{-0.05}$ & $0.99^{+0.12}_{-0.16}$
\\
~~~~S \dotfill 
 & $\cdots$ & $\cdots$ 
 & $1.12^{+0.08}_{-0.05}$ & $0.85^{+0.17}_{-0.22}$
 & $1.13^{+0.14}_{-0.18}$ & $1.07^{+0.23}_{-0.25}$
 & $1.13^{+0.13}_{-0.12}$ & $1.24^{+0.19}_{-0.27}$
\\
~~~~Fe \dotfill 
 & $\cdots$ & $\cdots$ 
 & $0.69^{+0.04}_{-0.05}$ & $0.61^{+0.04}_{-0.02}$
 & $0.68^{+0.09}_{-0.10}$ & $0.63^{+0.13}_{-0.10}$
 & $0.69^{+0.07}_{-0.08}$ & $0.78^{+0.16}_{-0.14}$
\enddata
\tablecomments{
The second and third columns show the fitting results of the absorbed 
\vmekal\ and \vnei\ models in XSPEC, respectively, (\vnei+\vmekal\ model 
is shown in Table~\ref{T:center_d}), for region ``C''.
In the last six columns, an absorbed recombining plasma model \neij\ in 
SPEX is applied for region ``C" and 5 smaller regions (see spectra in 
Figure~\ref{f:cspec_spex}).
Statistical errors of the fitted parameters are given at the 90\% 
confidence level, except the error ranges of the average charge and $kT_z$,
which are instead estimated using the 90\%\ confidence ranges of $kT$, 
$kT_i$, and $\tau$.
}
  \tablenotetext{a}{\phantom{0} $\tau$ represents the ionization
timescale for the \vnei\ model, while it is the recombination
timescale for the \neij\ model.}
  \tablenotetext{b}{\phantom{0} The tied abundances of C, N, O, Ne, Ar, 
Ca, and Ni.  }
  \tablenotetext{c}{\phantom{0} For the density estimates, we assume a 
sphere for region ``C" and short cylinders for all other rectangle regions.}
  \tablenotetext{d}{\phantom{0} The unabsorbed fluxes in the 
0.3--5.0 keV band.}
  \tablenotetext{e}{\phantom{0} The best-fit values of the initial 
electron temperatures all reach the upper limit of 10 keV and the error 
bars are poorly constrained.}
\label{T:center_s}
\end{deluxetable*}

\begin{figure*}[tbh!]
\centerline{ {\hfil\hfil
\psfig{figure=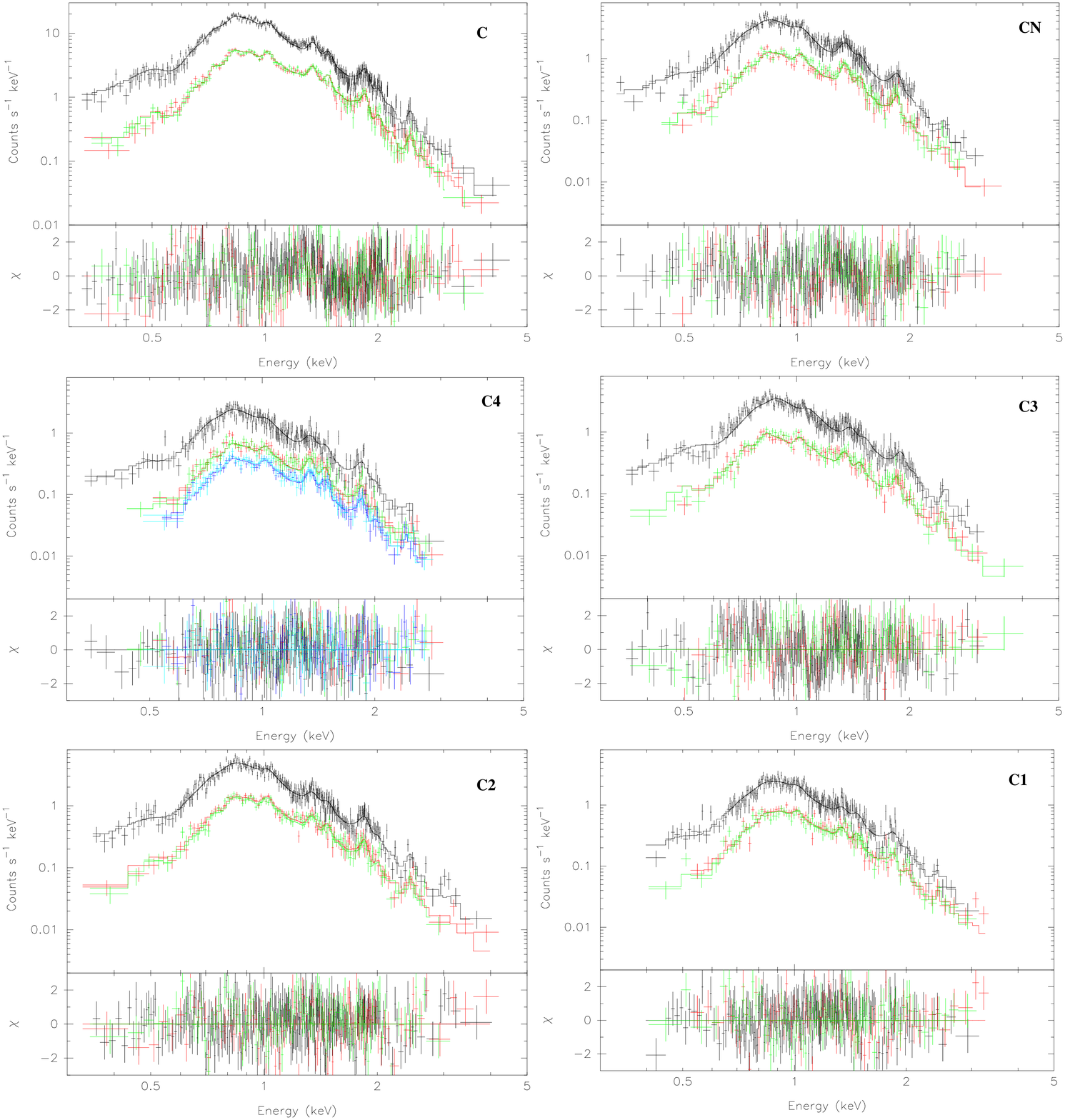,width=\textwidth,angle=0, clip=}
\hfil\hfil}}
\caption{
\XMMN\ EPIC spectra and fitted recombining plasma model from the 
six regions (indicated in Figure~\ref{f:reg}) in the SNR interior. 
Each individual spectrum is adaptively binned to achieve a 
background-subtracted S/N of 3.
The black solid lines show pn spectra, while MOS spectra are shown by 
the lower solid lines in red and green. The blue and cyan lines 
for region ``C4'' represent MOS1 spectra extracted from the southwestern 
edge of the NE observations with observation IDs 0145970101 and 0145970401, 
respectively.  
}
\label{f:cspec_spex}
\end{figure*}

We also examined whether a recombining plasma model could fit the 
\XMMN\ spectra in the six central regions, considering that there is
a weak bump at around 2.0~keV for the H-like Si line and that SK12 had 
adopted a recombining scenario to explain the plasma in the SNR interior.
We applied the NEI jump (\neij) model in SPEX,
which describes a CIE plasma with initial temperature $T_i$ rapidly
heated/cooled to a temperature $T$, observed after relaxation with 
a timescale $t$.
The fit results of the \neij\ model are tabulated in 
Table~\ref{T:center_s} and the spectra are shown in 
Figure~\ref{f:cspec_spex}.
In this recombining plasma model, the electrons of the X-ray-emitting
plasma in the region ``C'' with initial temperature $kT_i>6.5$ keV 
rapidly 
cool to a temperature of 0.6~keV after a timescale $t$ (the recombination
time scale $n_{\rm e} t\simeq 8.6 \E{11}~\cm^{-3}~\s$).
The foreground column density ($4.0\E{21}~\cm^{-2}$) and electron
temperature ($0.6$~keV) is similar to those from the single 
\vmekal\ and \vnei\ models, while the average charges of Mg, Si, S, 
and Fe (11.11, 12.65, 14.34, and 18.08, respectively) are larger 
than those expected from a 0.6-keV CIE plasma (10.71, 12.11, 13.95, 
and 17.22, respectively).
The estimated ionization temperatures of the four elements (0.7, 0.9, 
1.1, and 0.7~keV, respectively) are all higher than the 
electron temperature, indicating the four elements are over-ionized.
The most prominent residuals are shown as a bump at around 1.34~keV.
Adding a Gaussian line to compensate for the residuals,
we found that the bump is centered at $1.34^{+0.01}_{-0.02}$~keV and has
a width of $0.21^{+0.02}_{-0.04}$~keV ($\chi_\nu^2=1.24$).
Some of the residuals at $\sim 1.17$--1.29~keV may result from the incomplete
atomic data for Fe L-shell transition of SPEX code ($n=6,7,8\rightarrow 
2$ for \ion{Fe}{17}, $n=6,7\rightarrow 2$ for \ion{Fe}{18}, and 
$n=6\rightarrow 2$ for \ion{Fe}{19}; Brickhouse \etal\ 2000; 
Audard \etal\ 2001; also pointed out by SK12).
However, most residuals are above 1.3~keV and may result from the fit
itself rather than the incompleteness of the atomic data.

\begin{deluxetable*}{lcccccc}
\tabletypesize{\scriptsize}
\tablecaption{Spectral fitting results for the central gas of \snr\
with a two-component model}
\tablewidth{0pt}
\tablehead{
\colhead{Regions} & \colhead{C} & \colhead{CN} & \colhead{C4}
 & \colhead{C3} & \colhead{C2} & \colhead{C1}
}
\startdata
$\chi_{\nu}^{2}$ (d.o.f.) \dotfill
 & 1.32 (619) & 1.26 (453) & 1.04 (549) & 0.98 (437) & 1.18 (495) & 1.12 (391)\\
$\NH$ ($10^{21}\cm^{-2}$) \dotfill  
 & $5.50^{+0.45}_{-0.56}$ & $7.24^{+0.96}_{-1.01}$ & $5.91^{+0.84}_{-1.00}$
 & $3.86^{+0.35}_{-0.30}$ & $4.33^{+1.32}_{-0.20}$ & $3.57^{+0.50}_{-0.41}$\\
\hline
\colhead{Cold component} \\ 
$\kTc$ ($\keV$) \dotfill
 & $0.36^{+0.06}_{-0.04}$ & $0.28^{+0.05}_{-0.03}$ & $0.31^{+0.07}_{-0.05}$
 & $0.57^{+0.02}_{-0.05}$ & $0.54^{+0.04}_{-0.04}$ & $0.64^{+0.03}_{-0.03}$\\
$\tauc$ ($10^{11}\cm^{-3}\,{\rm s}$) \dotfill
 & $4.34^{+2.93}_{-1.73}$ & $5.14^{+12.04}_{-2.97}$ &$5.03 (> 1.72)$
 & $<500$ & $1.54^{+0.60}_{-0.70}$ & $7.28 (> 1.48)$\\
Abun$^{\rm a}$ \dotfill
 & $0.17^{+0.02}_{-0.03}$ & $0.30^{+0.11}_{-0.09}$ & $0.14^{+0.06}_{-0.05}$
 & $0.31^{+0.17}_{-0.15}$ & $0.15^{+0.06}_{-0.05}$ & $0.49^{+0.40}_{-0.32}$\\
${\rm [Mg/H]}$ \dotfill
 & $0.30^{+0.04}_{-0.03}$ & $0.75^{+0.21}_{-0.18}$ & $0.44^{+0.10}_{-0.09}$
 & $0.44^{+0.21}_{-0.15}$ & $0.37^{+0.08}_{-0.08}$ & $0.71^{+0.43}_{-0.45}$\\
${\rm [Si/H]}$ \dotfill
 & $0.24^{+0.03}_{-0.04}$ & $0.45^{+0.16}_{-0.12}$ & $0.34^{+0.09}_{-0.08}$
 & $0.37^{+0.15}_{-0.12}$ & $0.34^{+0.08}_{-0.08}$ & $0.44^{+0.25}_{-0.22}$\\
${\rm [S/H]}$ \dotfill
 & $0.34^{+0.06}_{-0.07}$ & $0.39^{+0.23}_{-0.19}$ & $0.53^{+0.22}_{-0.18}$
 & $0.52^{+0.23}_{-0.21}$ & $0.42^{+0.14}_{-0.14}$ & $0.32^{+0.29}_{-0.24}$\\
${\rm [Fe/H]}$ \dotfill
 & $0.19^{+0.03}_{-0.02}$ & $0.44^{+0.11}_{-0.13}$ & $0.27^{+0.06}_{-0.07}$
 & $0.25^{+0.05}_{-0.06}$ & $0.31^{+0.05}_{-0.05}$ & $0.40^{+0.18}_{-0.16}$\\
$f_{\rm c} n_{e,c} \nHc V\du^{-2}~(10^{57}\cm^{-3})$ \dotfill  
 & $7.78^{+2.86}_{-2.61}$ &$3.15^{+2.69}_{-1.55}$ &$1.61^{+1.68}_{-0.79}$
 & $0.56^{+0.21}_{-0.15}$ & $0.67^{+0.20}_{-0.12}$ & $0.30^{+0.22}_{-0.10}$\\
$\nHc \du^{1/2} (\cm^{-3})^{\rm b}$ \dotfill
 & 1.48 & 1.79 & 2.03 & 1.04 & 1.17 & 0.91 \\ 
$F_{\rm c}$~($10^{-11} \erg\cm^{-2}\ps$)$^{\rm c}$  \dotfill
 & 10.36 & 6.02 & 2.06 & 1.13 & 1.58 & 0.84 \\  \hline
\colhead{Hot component} \\
$\kTh$ ($\keV$) \dotfill
 & $0.77^{+0.02}_{-0.02}$ & $0.78^{+0.06}_{-0.04}$ & $0.78^{+0.04}_{-0.04}$
 & $1.20^{+0.21}_{-0.25}$ & $1.00^{+0.06}_{-0.06}$ & $1.40^{+0.19}_{-0.35}$\\
$f_{\rm h} n_{e,h} \nHh V\du^{-2}~(10^{57}\cm^{-3})$ \dotfill  
 & $3.09^{+0.26}_{-0.23}$ & $0.44^{+0.11}_{-0.09}$ & $0.40^{+0.07}_{-0.07}$
 & $0.13^{+0.10}_{-0.04}$ & $0.36^{+0.08}_{-0.06}$ & $0.14^{+0.08}_{-0.03}$\\
$f_{\rm h}^{\rm d} $ \dotfill
 & 0.65 & 0.52 & 0.61 & 0.52 & 0.65 & 0.69 \\
$\nHh \du^{1/2} (\cm^{-3})^{\rm b}$ \dotfill
 & 0.69 & 0.64 & 0.81 & 0.50 & 0.63 & 0.42 \\ 
$F_{\rm h}$~($10^{-11} \erg\cm^{-2}\ps$)$^{\rm c}$  \dotfill
 & 5.52 & 1.24 & 0.83 & 0.28 & 0.75& 0.32  
\enddata
\tablecomments{
An absorbed thermal model \vnei\ (cold component)+\vmekal\ (hot component) 
is used for region ``C" and 5 smaller regions 
(see spectra in Figure~\ref{f:cspec}).
The metal abundances of the cold and hot components are coupled.
Statistical errors are given at the 90\% confidence level.
}
  \tablenotetext{a}{\phantom{0} The tied abundances of C, N, O, Ne, Ar,
Ca, and Ni.}
  \tablenotetext{b}{\phantom{0} In the estimate of densities, we assume a 
sphere for region ``C" and short cylinders for all other rectangular regions.}
  \tablenotetext{c}{\phantom{0} The unabsorbed fluxes of the cold component
($F_{\rm c}$) and of the hot component ($F_{\rm h}$) are in the 0.3--5.0
keV band.}
  \tablenotetext{d}{\phantom{0} Volume filling factor of the 
hot-phase gas $f_{\rm h}=1-f_{\rm c}$, where the filling factor of 
the cold-phase gas $f_{\rm c}=(1+(\Th/\Tc)^2({\rm norm_{h}/
norm_{\rm c}}))^{-1}$.}
\label{T:center_d}
\end{deluxetable*}

We further applied the recombining plasma model \neij\ to the five 
smaller regions.
As shown in Table~\ref{T:center_s}, the model gives a smaller variation 
of temperature (0.54--0.63~keV) and column density (3.7--$4.3\E{21}
\cm^{-2}$) than given by a \vnei+\vmekal\ model.
Compared to the two-temperature model, the recombining plasma model 
gives similar $\chi_\nu^2$ value for region ``C2'' and a marginally 
poorer fit for regions ``C1'' and ``C4''.
Hence, we cannot exclude either of the two-temperature or the 
recombining plasma model for the three smaller scale regions with 
the current available \XMMN\ observations.
However, for regions ``CN'' and ``C3'', the $\chi_\nu^2$ values of the 
\neij\ spectral fits are apparently larger than those of the 
two-temperature model (1.44 and 1.35 versus 1.26 and 0.98, respectively).
We added a Gaussian line for regions ``CN'' and ``C3'' to compensate 
for the residuals at around 1.3 keV and found that the fits 
($\chi_\nu^2=1.36$ and 1.28, respectively) are still poorer than the two-temperature 
model, which implies that the latter model provides a better fit to 
the data for the two regions.

\section{Discussion} \label{S:discussion}
\subsection{Global Gas Properties} \label{S:global}
We have performed an \XMMN\ imaging and spatially-resolved spectroscopic 
study of the MM~SNR \snr.
In the X-ray band, the remnant shows a centrally peaked morphology and
blobby structures in the SNR interior.
The limb-brightened shell in the northeast appears distorted by the 
impact with dense ambient medium.

\subsubsection{Gas Properties of the NE Shell} \label{S:NEgas}
The X-ray emission from the NE limb is best described by a combination 
of a $kT=0.3$~keV thermal component with sub-solar metal
abundances and a hard component of either thermal or non-thermal
origin (see Section~\ref{S:NEresult} for details).
We investigated the possible spatial variation of the gas
properties by dividing the bright NE shell into two smaller regions
``S1'' and ``S2''.  No apparent variations have been found between them.

According to the \vnei+\powerlaw\ model, the density of the X-ray-emitting gas
in region ``Shell'' can be estimated from the volume emission measure
$f \nel \nH V =4.4\pm0.6 \E{57}\du^2~\cm^{-3}$, where $f$ 
is the filling factor of the X-ray-emitting plasma within the volume $V$,
$\nel$ and $\nH$ are the electron and hydrogen density of the gas, 
respectively, and $\du=d/2~\kpc$ is the distance scaled with 2~kpc.
Assuming an oblate spheroid for the elliptical region ``Shell" (with 
half-axes $3\farcm{68} \times 3\farcm{68} \times 1\farcm{56}$) 
and $\nel=1.2 \nH$, the mean hydrogen density and the mass of 
the X-ray emitting gas are
obtained to be $\nH=2.7\pm0.2 f^{-1/2}\du^{-1/2}~\cm^{-3}$ and 
$M{\rm_X} =1.6\pm0.1 f^{1/2} \du^{5/2}~\Msun$, respectively.
The ambient density is thus estimated to be $n_0=\nH/4\sim0.7 f^{-1/2} 
\du^{-1/2}~\cm^{-3}$.
The ionization age inferred from the ionization timescale $n_e t_i$ 
is $t_{\rm i}> 7.5 f^{1/2}\du^{1/2}$~kyr.

The two-temperature (0.3~keV+0.6~keV) model can also represent the spectra
of the NE shell.
We assume that the two-phase gas fills the whole volume (with filling factors
$f_{\rm c}+f_{\rm h} =1$) and is in pressure equilibrium.
The colder gas is found to fill 68\% of the volume, with a density
$\nHc=2.5\pm 0.3 (f_{\rm c}/0.68)^{-1/2}\du^{-1/2}~\cm^{-3}$, while the hotter
gas has a density $\nHh=1.2\pm0.2 (f_{\rm h}/0.32)^{-1/2}\du^{-1/2}~\cm^{-3}$.
The mass of the cold and hot components is estimated to be
$\sim 1.0 (f_{\rm c}/0.68)^{1/2} \du^{5/2}~\Msun$ and
$\sim 0.2 (f_{\rm h}/0.32)^{1/2} \du^{5/2}~\Msun$, respectively.
As the X-ray emission is emitted from the shock--cloud interaction 
region, 
the hot plasma may consist of two density components: (1) the dense 
cloud material and (2) the tenuous inter-cloud medium. 
The denser cloud material heated by the transmitted shock is relatively
colder, which is responsible for the 0.3 keV component, while the
inter-cloud medium may give arise to the $\sim 0.6$~keV component, as 
also seen in the shells of the Vela SNR (Bocchino \etal\ 1999; 
Miceli \etal\ 2006) and the Cygnus Loop (Zhou \etal\ 2010).
The ionization timescale (0.5--$1.6 \E{12}~\cm^{-3} \s$) of the hot 
inter-cloud gas can provide a rough estimation of its ionization age 
of 1--$4\E{4}~\yr$.
Therefore, both of the two-temperature and \vnei+\powerlaw\ models suggest
an evolved stage for \snr.

\subsubsection{Recombining Plasma in the Interior of \snr?} \label{S:dis_rec}
We found that a recombining plasma model for over-ionized gas 
could fit the spectra of the central region, although the model is 
neither the only plausible one nor the best-fit one when compared 
with the two-temperature model (see Section~{\ref{S:result_cent}).
In this model, the central X-rays arise from a recombining plasma with
an electron temperature $kT=0.59$~keV, and multiple ionization 
temperatures of elements increasing from Mg ($\kTz=0.72$~keV) to S 
($\kTz=1.12$~keV) and dropping at Fe ($\kTz=0.69$~keV).
Another multi-$\kTz$ model is previously used by SK12, although 
different $\kTz$ and $kT$ values were obtained (see further discussion 
in Section~\ref{S:comp}).
The monotonic rise of the ionization temperature with atomic number except
for Fe was explained by the element-dependent recombination timescale
(SK12; Smith \&  Hughes 2010).
Assuming the filling factor of the X-ray-emitting gas is close to 1,
the hydrogen density of the gas is estimated as $\nH\sim0.8
\du^{-1/2}~\cm^{-3}$.
The recombination age is inferred from the recombination timescale 
$t_{\rm rec}=\tau/n_e\sim 2.9\E{4}\du^{1/2}~\yr$, slightly lower than 
but consistent with the dynamical age of \snr\ (3--$4 \E{4}~\yr$).

\begin{deluxetable*}{p{3.3cm}ccccc}
\tabletypesize{\footnotesize}
\tablecaption{Comparing our \XMMN\ spectral results for the central 
gas with previous X-ray studies }
\tablewidth{0pt}
\tablehead{
 \colhead{Telescope} & \multicolumn{2}{c}{\XMMN}  & 
 \colhead{\ASCA/\ROSAT} & \colhead{\ASCA} & \colhead{\Suzaku} \\ 
\colhead{Paper} & \multicolumn{2}{c}{Present}  & \colhead{RB02} &
\colhead{KO05$^{\rm a}$} & \colhead{SK12} \\
 \cline{2-3}
 \colhead{Model} & \vnei+\vmekal & \neij & \vnei+\vnei & \vnei+\vnei
 & Multi-$T_z$ Recombining
}
\startdata
$\chi_{\nu}^{2}$ (dof) \dotfill
 & 1.32 (619) & 1.44 (620) & 1.1 ($\sim 650$) & 1.06 (129)& 1.41 (634) \\
$\NH$ ($10^{21}\cm^{-2}$) \dotfill  
 & $5.5^{+0.4}_{-0.6}$ & $4.0 \pm 0.1$  & $6.6 \pm 0.6 $
 & $5.1^{+0.8}_{-0.2}$ & $4.7^{+0.2}_{-0.1}$ \\
$\kTc$ ($\keV$) \dotfill
 & $0.36^{+0.06}_{-0.04}$ & $0.59 \pm 0.01$ & $0.59 \pm 0.05$ 
 & $0.62 \pm 0.03$ & $0.40^{+0.02}_{-0.03}$ \\
$\tauc^{\rm b}$ ($10^{12}\cm^{-3}\,{\rm s}$) \dotfill
 & $0.43^{+0.29}_{-0.17}$ & $0.86^{+0.06}_{-0.09}$ & $2 (> 1)$
 & $ 31 (>2.7)$ & $0.40 (>0.32)^{\rm c}$\\
$\kTh$ ($\keV$) \dotfill
 & $0.77^{+0.02}_{-0.02}$ & $\cdots$
 & $1.8^{+0.7}_{-0.5}$ & $1.41^{+0.26}_{-0.10}$ & $\cdots$ \\
$\tauh$ ($10^{12}\cm^{-3}\,{\rm s}$) \dotfill
 & $\cdots$ & $\cdots$ 
 & $0.5^{+0.5}_{-0.4}$ & $0.5^{+1.6}_{-0.2}$ &  $\cdots$
\enddata
\tablecomments{Different source region (except current study and RB02 
used similar source region ``C") and background region are selected 
for spectral extractions in each study.
}
  \tablenotetext{a}{\phantom{0} The best-fit parameters for the central 
region indicated by the black rectangle shown in Figure~\ref{f:reg}. }
  \tablenotetext{b}{\phantom{0}
$\tauc$ represents the ionization
timescale for the cold \vnei\ component, while it is the recombination
timescale for the multi-$T_z$ model.}
  \tablenotetext{c}{\phantom{0} See SK12 for the detailed calculation 
of the recombination timescale.}
\label{T:compare}
\end{deluxetable*}

By applying the recombining plasma model to the five smaller regions, 
we found a small spatial variation of the parameters.
The electron temperature reaches a maximum in the western region ``C1''
($\sim0.63$ keV) and a minimum at region ``C3'' ($\sim 0.54$ keV).
The recombination ages of regions ``C1'', ``C2'', ``C3'', and ``C4''
are estimated as $2.3\E{4}$ yr, $2.4\E{4}$ yr, $2.1\E{4}$ yr, 
and $2.5\E{4}$ yr, respectively, while the region ``CN'',
which is basically along the inner radio shell,
shows the longest recombination age ($\sim 5.3\E{4}$ yr) 
with the lowest metal ionization temperatures (that are close to the 
electron temperature $kT\sim 0.6$ keV).

\subsubsection{Two-temperature Gas in the SNR Center} \label{S:twoT}
The X-ray spectra of the gas in the central regions are best
represented by a two-temperature model including an under-ionized 
cold phase and a CIE hot phase (see Table~\ref{T:center_d}).
We have not found any evidence of ejecta inside the SNR because all 
metal abundances are below the solar value.
Under the assumption that the cold gas and hot gas in region ``C'' are in 
pressure equilibrium ($\nHc \Tc=\nHh \Th$) and fill all the volume 
($f_{\rm c} + f_{\rm h}=1$), we obtain the filling factor of the two 
components as $f_{\rm c}\sim 0.35$ and $f_{\rm h}\sim0.65$, with 
corresponding mean hydrogen densities $\nHc\sim1.5 
\du^{-1/2}~\cm^{-3}$ and $\nHh \sim0.7 \du^{-1/2}~\cm^{-3}$.
The ionization age of the cold-phase gas is inferred to be around 7.7 
kyr, shorter than the dynamical age of the SNR, implying the gas could be
more recently shocked or evaporated.
We also estimate the gas densities of the smaller, rectangular regions
in the SNR interior by assuming short cylinders as the three-dimensional 
shapes. The derived parameters, including the densities, filling 
factors, unabsorbed fluxes of the two-phase gas in the five regions, 
are tabulated in Table~\ref{T:center_d}.

The spatially resolved spectroscopic analysis allows us to find the
variations of the physical properties.
There is a large-scale column density gradient that decreases from the
northeast to the southwest.
The heaviest absorption occurs on the NE shell (``Shell'':
$\NH\sim 8\E{21}~\cm^{-2}$) and the inner shell (``CN'':
$\NH\sim 7\E{21}~\cm^{-2}$), while the lowest one is found in the west 
of the remnant (``C1'': $\NH\sim4\E{21}~\cm^{-2}$).
The large-scale absorption distribution is consistent with the optical 
reddening dropping from the northeast to the X-ray center (Long 
\etal\ 1991), which both agree with the picture that considerable 
molecular gas is located in the northeast.
The enhanced absorption on the inner radio shell also favors that
the inner shell is a shock--MC interaction interface projected 
to the northern hemisphere (Dubner \etal\ 2000).

We also found a temperature variation inside \snr , which generally 
shows an opposite trend against the absorption.
The temperature of the cold component drops by a factor two
from the western region (``C1") to the inner radio shell 
(``CN") and in the east (``C4");
in the latter regions, the gas density of the cold component
is higher than that in the west.
Such variations of the foreground absorption and the temperature 
and density of the cold component are reasonable for an SNR evolving 
within, and mixing with, a denser medium (especially molecular gas) 
in the north and east.

Furthermore, two distinct \Ha\ patterns, filamentary and diffuse, are 
present in the denser regions (``CN" and ``C4") and the tenuous regions 
(``C1", ``C2" and ``C3"; as shown in Figure~\ref{f:ximg}(b)),
respectively.
The [\SII]/\Ha\ ratios on the inner radio shell and on the 
NE limb were relatively high, ranging from 0.4 to 1.2, favoring that 
shock ionization occurs on the two shells and the \Ha\ emission arises 
from a post-shock recombination zone (Long \etal\ 1991), whereas the 
SNR center has a low [\SII]/\Ha\ ratio (Keohane \etal\ 2005; Figure~3).
The distinct \Ha\ patterns and [\SII]/\Ha-ratios in the SNR interior 
compared to those along the shells imply that the optical emission 
in the hot remnant interior has a different origin from that on the 
shells (see more in Section~\ref{S:interp}).

\subsection{Comparision with the \ROSAT, \ASCA\ and \Suzaku\ Results} 
\label{S:comp}

The X-ray spectra of the SNR interior were previously investigated by 
RB02 with \ROSAT\ and \ASCA, by KO05 with \ASCA\ and by SK12 with 
\Suzaku\ and some of their spectral results are summarized in 
Table~\ref{T:compare}.
The \ROSAT\ and \ASCA\ observations cover the whole X-ray-emitting 
area, while the \Suzaku\ study was toward the SNR interior.
Different source (see Figure~\ref{f:reg}) and background regions are 
selected for spectral analysis in those studies.
The background spectra of RB02 and  KO05 are subtracted from the blank 
sky observations and a nearby source-free region (without specifying the 
location), respectively, while SK12 subtracted the background from a 
sky field ($\sim 1\fdg{5}$ away) on the Galactic plane, respectively.
In our study here, we improved on the background subtraction by 
adequately selecting a region as close as possible to the observation
(only $\sim30 '$ away) and with the same Galactic latitude, thus 
minimizing any contamination by the Galactic background emission.
We here compare the plasma properties determined from our \XMMN\ 
analysis with those from the previous studies.

In the studies made by RB02 and KO05, the central X-ray-emitting gas 
is characterized by a two-component NEI model with similar $\kTc$ 
($\sim 0.6$ keV) and relatively high $\kTh$ ($>1.3$ keV). 
Nevertheless, the two studies gave inconsistent interstellar absorption 
and metal abundances.
KO05 also extracted a spectrum in the east of their ``center'' region
and found a smaller ratio of high-$T$ emission measure to the low-$T$ 
emission measure ($\sim0.3$) in the eastern region than that in 
the center ($\sim0.5$).
The emission measure ratios of hot-to-cold gas in our \XMMN\ study 
($\sim 0.5$ in the regions ``C1'' and ``C2''; $\sim 0.3$ in the area 
containing ``C2'', ``C3'' and ``C4'') are similar to those obtained 
by KO05.
Unlike the \ASCA\ and \ROSAT\ studies, our \XMMN\ analysis based on 
a two-temperature model gives different values of the temperature and 
the ionization timescale, although we select a region (``C'') 
in the SNR center similar to that RB02.

A discrepancy was also previously pointed out by SK12 by comparing 
their \Suzaku\ study with the results provided by RB02:
the \Suzaku\ spectrum does not reveal an Fe \Ka\ line and the central 
gas has much lower temperature values ($\kTc=0.24$~keV and $\kTh=$
0.77~keV) when a two-temperature CIE model is used to fit the spectrum.
For the \Suzaku\ study, a multi-$T_{\rm z}$ recombining plasma model 
was preferred to describe the central gas, in order to better 
fit the line-like and bump-like residuals at the Si \Lya\ energy and 
in the 2.4--5.0 keV range.
As noted earlier, the \Suzaku\ background was carefully selected on 
the Galactic plane $\sim 1\fdg{5}$ away, and thus SK12 argued that a 
better selection of background brought the distinguished spectroscopic 
results, while the \ASCA\ study of KO05, which also selected a nearby 
background, found no evidence of over-ionization.
In our \XMMN\ study, we have shown that the multi-$T_z$ recombining
plasma model is feasible, although not the best-fit model to explain
the spectra of the central gas.
However, we obtained higher electron and ionization temperatures and 
larger recombination timescale compared to those provided by SK12.

The spatial variations of the physical parameters found in this study
indicate that a different background selection, in addition
to the different spatial and spectral capabilities of the telescopes
used, are the likely origin for the inconsistent results obtained by 
the four X-ray studies.

Furthermore, the blobby central gas as revealed by the \XMMN\ and 
\Chandra\ imaging studies indicates that the X-ray emission may be 
contributed by more than one component, and our two-temperature 
thermal component model seems to be a natural explanation for this.
Confirming the RRC and the recombining plasma model for the central
thermal X-ray emission will require high-resolution spectra which 
can be obtained with the upcoming mission {\it ASTRO-H} (Takahashi
\etal\ 2012).

\subsection{Non-thermal Emission from the NE Shell?}}

In the northeast, the shock is propagating into dense MCs, exciting
1720 MHz OH masers, and simultaneously generating VHE \rray s, 
making the remnant one of the best sites for comic ray 
acceleration studies.
As mentioned in Section~\ref{S:NEresult}, an absorbed thermal+non-thermal
model can reproduce the X-ray spectra of the NE shell, 
although the two-thermal component model is also feasible and can not be 
ruled out by using the currently available \XMMN\ data.
Here, we discuss the hard non-thermal component based on the 
\vnei+\powerlaw\ model.

The hard X-ray tail in spectra of the NE shell is best fitted with 
a \powerlaw\ model with a photon index $\Gamma =0.9$--2.4.
The unabsorbed flux of this non-thermal emission is $F{\rm ^{NT}_X} 
\sim 3.4 \E{-13}~\erg \cm^{-2} \ps$ (0.3--5.0 keV), corresponding to a 
luminosity $L{\rm ^{NT}_X} \sim 1.6 \E{32} \du^2~\erg \ps$.
There are three potential origins for the non-thermal X-rays:
synchrotron emission, inverse Compton (IC) scattering, and 
bremsstrahlung from non-thermal electrons.

X-ray synchrotron emission has been generally detected in the fast-moving
($\gsim 2000$ \kms) blast waves of historical SNRs such as Cas A, 
{\it Tycho}, {\it Kepler}, and SN1006 (see Ballet 2006, and references 
therein).
The electrons emitting synchrotron X-rays in the keV band requires 
energies over tens of TeV in a typical magnetic field ($B\approx 
10$--$500\mu$G; Vink 2012).
The shock velocity in \snr\ measured from the optical observations is 
60--100 \kms\ (Bohigas \etal\ 1983; Long \etal\ 1991), which is too slow 
to effectively accelerate the electrons to such high energies.
The cut-off energy of the synchrotron emission for the 
shock velocity $V_{\rm s}\sim 100 \km\ps$ is 
$h \nu_{\rm cut-off}=0.56\eta^{-1} [(\chi_4-1/4)/\chi_4^2] (V_{\rm s}/100 \km
\ps)^2$~eV ($\chi_4$ is the compression ratio in units of 4, the deviation
from Bohm diffusion $\eta\lsim 10$; Vink 2012),
which is not in the X-ray band.
The synchrotron spectrum above the cut-off photon energy 
is thus rather steep in the X-ray band because of the quick roll over of
the spectrum.
Hence, the synchrotron radiation is not expected to be the dominant 
scenario due to the acceleration difficulty and hard spectrum.
IC scattering is usually not considered to be important below 10
keV (Vink 2012), and is therefore also discarded here as the origin
of the non-thermal X-ray emission.

\subsubsection{Non-thermal Bremsstrahlung} \label{S:brem}
Bremsstrahlung in the X-ray regime can be produced by suprathermal 
electrons of energy below several MeV.
At these electron energies, Coulomb collisions dominate over other 
loss mechanisms, and contribute to the flattening of the spectrum 
and to lowering the observed flux (Sturner \etal\ 1997).
The loss-flattened spectrum yields a hard photon index $\Gamma \approx 
s-1$, where $s$ is the spectral index of the parent electrons.
Uchiyama \etal\ (2002) utilized this scenario to explain the extremely 
flat X-ray spectrum (with a photon index of $\Gamma \approx 0.8$--1.5) 
of several clumps in the evolved SNR $\gamma$-Cygni.
A similar scenario can be applied to the case of the NE part of \snr\
($\Gamma=0.9$--2.4).

In \snr, the non-thermal bremsstrahlung emitted by electrons suffering 
Coulomb losses is a potential explanation for the faint, relatively flat 
X-ray emission from the NE shell. 
The non-thermal electrons can be accelerated by radiative shock 
waves propagating into MCs (Bykov \etal\ 2000).
If non-thermal bremsstrahlung is the dominant contributor to the hard 
X-ray continuum detected in the dense shell, we expect to obtain 
a spectral index of $s=1.9$--3.4 for the parent elections.

\subsubsection{Secondary Leptons}

\begin{figure}[t]
\centerline{ {\hfil\hfil
\psfig{figure=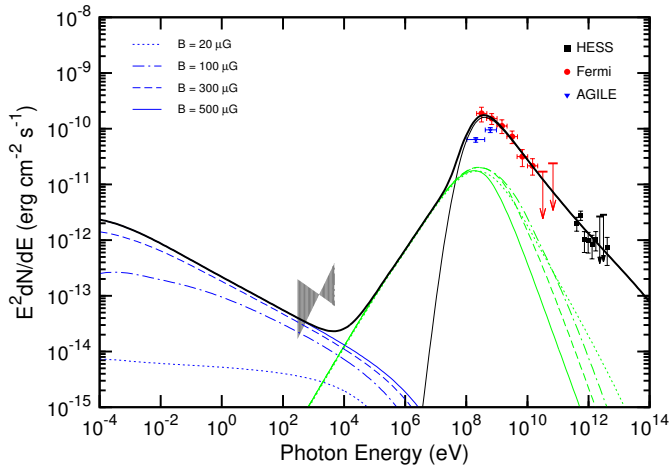,width=3.7in,angle=0, clip=}
\hfil\hfil}}
\caption{
Broad-band energy spectrum (thick solid line) for a molecular cloud of 
mass $5\E{4} \Msun$.
The observational \rray\ and non-thermal X-ray emission are indicated by 
dots (with error bars) and a gray bow-tie, respectively.
The black solid line shows the emission from $\pi^{0}$-decay. 
Synchrotron emission and bremsstrahlung from secondary electrons 
are colored in blue and green, respectively. 
The dotted, dash-dotted, dashed, and thin solid lines refer to the 
emission for magnetic field values of $B_{\rm cl}$ of $20\mu G$, 
$100 \mu$G, $300 \mu$G and $500 \mu$G, respectively.
}
\label{f:sec}
\end{figure}

The hadronic origin of the \rray\ emission in \snr\, has been 
widely discussed and accepted.
Accompanying the $\pi^{0}$-decay $\gamma$-rays, secondary electrons 
from the decay of the charged pions can radiate non-thermal emission
through the synchrotron and bremsstrahlung processes.
The broadband non-thermal emission from the MCs illuminated by CRs 
accelerated in SNRs has been modeled by Gabici \etal\ (2009). 
We adopt this model here for \snr\ to calculate the broadband spectrum 
of the secondary electrons for the NE MCs (see Figure~\ref{f:sec}).
The following parameters for \snr\ have been adopted in this model:
age $t_{\rm age}=4\E{4}$~yr, supernova explosion energy $E_{\rm SN}=
6.6\E{50}$~erg, fractional energy deposited to CRs $\eta$ = 0.1, 
and distance of the MC to the remnant center $d_{\rm cl}\approx10$~pc 
(Li \& Chen 2010).
The shape of the spectrum is mainly determined by four parameters: the
correction factor of the diffusion coefficient ($\chi$), the power-law 
index of diffusion coefficient ($\delta$), the magnetic field in the MC 
($B_{\rm cl}$), and the cloud mass ($M_{\rm cl}$).
We initially determine the distribution of the secondary electrons by 
fitting the \rray\ data from the {\it Fermi} (Abdo \etal\ 2010) and H.E.S.S. 
(Aharonian \etal\ 2008) observations.
The fitting gives $\chi=0.08$,  $\delta=0.45$ and $M_{\rm cl} = 5\E{4} 
M_{\odot}$ which are consistent with the previous results obtained by Li 
\& Chen (2010).

The synchrotron radiation from the secondary electrons strongly depends 
on the strength of the magnetic field in the MC and gets enhanced with
increasing magnetic field values.
However, even if the magnetic field is increased up to $500~\mu$G, the 
synchrotron emission is not strong enough to explain the observed 
non-thermal X-rays.
Moreover, the bremsstrahlung from the secondary electrons, which peaks 
around 100~MeV, is also too low to fit the observed X-ray data.
The emissivity of bremsstrahlung from the secondary electrons is 
proportional to the gas density squared but not sensitive to the magnetic 
field.  Increasing the gas density, however, does not enhance the flux 
due to the electrons' strong energy loss by bremsstrahlung.
The spectrum extraction region ``Shell" in the NE edge just
overlaps a small part of the illuminated MC; 
however, the observed non-thermal X-ray flux is still much higher than 
could be produced by the secondary electrons from the whole MC. 
In summary, the secondary electrons' origin is not a favorable scenario 
to explain the hard X-ray tail.

\subsubsection{Radio Emission and Non-thermal Bremsstrahlung X-rays}

It is of interest to search for the relationship between the non-thermal 
X-rays and the radio synchrotron emission, since the bright X-ray shell 
appears spatially coincides with the radio peak.
It is suggested that, in an inhomogeneous medium, the fast shock
propagating in the intercloud medium accelerates particles to higher
energies and generates radio synchrotron and \rray\ emission, 
while the shock propagating in the clouds is able 
to accelerate electrons to energies below GeV giving birth to 
non-thermal hard X-ray and MeV \rray\ emission (Bykov \etal\  2000;
Chevalier 1999).

The overall radio spectral index of \snr\ is observed to be $\alpha
\sim 0.35$ ($S \propto \nu ^{-\alpha}$), with the index values on the 
bright radio filaments systematically flatter than in the SNR cavity 
(Dubner \etal\ 2000).
Accordingly, we infer the energy spectral index of the radio emitting 
electrons on the shell to be $s < 1.7$, which is much lower than the 
index of the parent electrons radiating bremsstrahlung X-rays 
($s=1.9$--3.4; see Section~\ref{S:brem} above).

This discrepancy implies that non-thermal radio and X-ray emission do 
not arise from the same population of electrons, and suggest that 
the radio emission likely originates from the intercloud shock while
the flat X-ray emission is associated with the cloud shock.

\subsection{The SNR Interior Gas} \label{S:interp}
As discussed in Sections~\ref{S:global} and \ref{S:comp}, 
the gas in the interior of \snr\ is spatially variant, clumpy, 
X-ray bright, and best characterized by a two-temperature plasma,
or even a recombining plasma.
In Sections~\ref{S:in_twoT} and \ref{S:in_rec},
we discuss the physics of the centrally bright X-ray emission
based on the former and latter models, respectively.

\subsubsection{Physics of the Central Thermal X-Ray Emission} \label{S:in_twoT}

Projection effect is one of the mechanisms that is suggested to
explain the centrally peaked X-ray morphology. 
One projection case occurs when the interaction zone between the SNR 
and a one-sided dense medium is projected, along the line of sight, 
to the remnant interior (Petruk 2001).
The enhanced intervening hydrogen absorption, high [\SII]/H$\alpha$ 
ratio, low temperature, and relatively high density
along the northern inner shell (typified by region ``CN")
indeed imply that a shock interaction with dense interstellar medium
(most probably, MC) is projectively seen in the interior.
Although the X-ray brightness seems to peak inside the inner
shell, we do not attribute it to the projection effect;
the reason is that the scenario that the bright X-rays come
from the shocked dense gas in the remnant's edge is inconsistent with 
the low [\SII]/\Ha\ ratio near the X-ray brightness peak.
Another type of projection effect was studied by Shimizu \etal\
(2012): 
When the shock breaks out of disk-like circumstellar medium to
a rarefied interstellar medium, the hot over-ionized plasma
appears to be centrally-filled (bar-like) in the equatorial plane.
The scenario was applied to describe the X-ray morphology
of the much younger SNR W49B (1000--4000~yr; see Shimizu \etal\
2012, and references therein; see also Zhou \etal\ 2011).
However, at the late-phase ($>10,000$~yr), the SNR evolves to be 
shell-like independently of the viewing angle rather than 
centrally-filled,
and \snr, at a large age (3--$4\E{4}~\yr$), would be of this case.
Hence, the projection effects are unlikely to explain the central
X-ray emission.

The thermal conduction model is often used to explain the central
X-ray emission of MM~SNRs (Cox et al.\ 1999; Shelton et al.\ 1999).
Without the suppression of magnetic fields, the conduction timescale
is given by $t_{\rm cond}\approx k\nel l_T^2/\kappa\sim 56(\nel/1~\cm^{-3})
(l_T /10 \pc)$ $(kT/0.6 \keV)^{-5/2} (\ln\Lambda/32)$~kyr, 
where $l_T$ ($=T_e/|\nabla T_e|$) is the scale length of the temperature 
gradient, $\kappa$ is the collisional conductivity, and $\ln\Lambda$ 
is the Coulomb logarithm.
If $l_T$ is taken as the radius of the SNR $\sim 10$~pc in the northern
hemisphere, and we use the temperature 0.77 keV and density 
$0.69~\cm^{-3}$ of the hot component in the central region ``C'', 
the conduction timescale is estimated to be 21 kyr, which is smaller 
than the dynamical age of \snr.
Hence, thermal conduction may play a role in the center of \snr.  
However, the almost monotonic decrease of the temperature of both the
hot and cold components from west to east across the central
region (see Section~\ref{S:twoT} and Table~\ref{T:center_d}) is 
inconsistent with the prediction of the thermal conduction model. 
Actually, RB02 have also regarded a large temperature gradient
(observed with \ROSAT\ and \ASCA) from the western ear-like 
structure to the eastern shell as a difficulty for the model 
to be applied to \snr.
Moreover, in the thermal conduction model, the pressure at the SNR
center is $\sim0.3$ of that at the shell and the central density
is much lower ($\sim 0.13$ at radiative shell formation) than the 
ambient gas density;
but in this study, the pressure ratio between the central gas 
and the NE shell is $\sim0.7$ and the density ratio is $\gsim 0.26$,
which are much larger than the model prediction.
These inconsistencies may be due to the over-simplified conduction 
model in comparison with a more complicated, highly nonuniform, 
interstellar environment and physical conditions.
Also, the magnetic field may be another potential factor to
suppress the effect of thermal conduction.

An alternative scenario for the centrally brightened X-ray
morphology is the cloudlet evaporation model (White \& Long 1991).
A series of observational facts in W28 seem to favor this scenario.
Firstly, blobby X-rays and diffuse \Ha\ emission are detected in the SNR 
center, indicating that the central gas is highly clumpy and thus 
probably contains plenty of cloudlets required by the model.
Secondly, the cloudlets embedded in a previously shocked, hot gas
are gradually evaporated, and the newly evaporated material is
expected to be dense and characterized by a low temperature plasma.
This can naturally explain the two-temperature component model of
the X-ray emission from the \snr\ interior.
Thirdly, the cold component in our best fit spectral model has 
a low ionization timescale ($\sim 4\E{11}~\cm^{-3}\s$) indicating 
an NEI state, and this is expected by the cloudlet evaporation model.
Moreover, the cloudlet evaporation model predicts that collisionally
excited \Ha\ emission is generated in the evaporating flow and has a
larger luminosity than the X-ray luminosity for evaporation-dominated 
SNRs, and this has indeed been pointed out to be the case in W28 
(Long \etal\ 1991).

Comparatively, the cloudlet evaporation mechanism can essentially
explain the properties of the X-ray emission in the central
region of W28, but the thermal conduction mechanism can also
play a role in a length scale comparable to the remnant's radius.
However, the limited field of view in the \XMMN\ observations and the
complicated X-ray brightness profile that RB02 obtained using
\ROSAT\ observation prevent us from a detailed quantitative
calculation using the evaporation model.

\subsubsection{Relation to the Recombining Plasma?} \label{S:in_rec}

The recombining plasma model is also a feasible model for the central
gas (see Section~\ref{S:dis_rec}). 
The presence of the over-ionized plasma needs a rapid cooling 
process,
such as drastic adiabatic expansion (including rarefaction process, Itoh 
\& Masai 1989), thermal conduction 
(Kawasaki \etal\ 2002) and both processes (Zhou \etal\ 2011).
The rarefaction scenario predicts a rapid electron cooling after the
shock breaking out of the dense circumstellar medium and adiabatically
expands to a tenuous ambient medium.
Though, the slight density variation inferred from the \XMMN\ analysis seems
not support a distinctive density decrease from the explosion center.
Thermal conduction to low-temperature gas in small clouds may also
rapidly cool the nearby hot gas.
As diffuse \Ha\ emission has been detected in the SNR center and
overlaps the hot X-ray-emitting plasma, it can be expected that 
many dense clouds emitting the optical emission 
are embedded in the hot plasma.
On the other hand, if cloudlet evaporation is not negligible,
it is very possible that the hot plasma could be rapidly cooled down 
by mixing with the dense and cooler plasma evaporated from the clouds.
Indeed, a numerical simulation by Zhou \etal\ (2011) has shown that the mixing 
of the gas evaporated from dense material is one important cooling 
mechanism to produce the over-ionized plasma in SNR W49B.

\section{Summary} \label{S:summary}
We have investigated the hot gas in the northern hemisphere of the MM~SNR 
\snr\ by performing an \XMMN\ imaging and spatially resolved 
spectroscopic study.  The main results and conclusions are summarized 
as follows:
\begin{enumerate}

\item The \XMMN\ image reveals a deformed shell in the northeast which 
partly overlaps the molecular gas and \Ha\ emission.
The X-ray emission in the interior is blobby and appears to be 
confined in the north and east by the \Ha\ filaments.

\item The X-rays arising from the NE shell, where the shock--MC 
interaction is evident and \rray\ emission is seen, consist of 
a soft thermal component ($\sim 0.3$ keV) plus 
a hard component of either non-thermal ($\Gamma=0.9$--2.4) 
or thermal ($\kTh\sim 0.6$~keV) origin.
The ionization age is estimated to be $> 0.75\E{4}$~yr and 1--$4\E{4}$~yr,
respectively, from the former model (\vnei+\powerlaw) and the hot
component of the latter model (double-\vnei), further supporting an
evolved stage of SNR \snr.
According to the \vnei+\powerlaw\ model, we estimate a density of 
$\sim 2.7\cm^{-3}$ for the hot post-shock plasma.
In the double-\vnei\ model, the soft X-ray component 
may originate in the shocked cloud material with a density of $\sim 2.5
\cm^{-3}$, while the harder X-rays may be emitted from the inter-cloud gas
with a density of $\sim 1.2\cm^{-3}$.

\item
If the hard X-ray emission of the NE shell is indeed non-thermal,
a possible origin for its flat spectrum is the non-thermal bremsstrahlung
emitted by the electrons suffering Coulomb losses.
The low spectral index and the difficulty in
accelerating electrons to $>$TeV-energy disfavor a synchrotron origin.
The hard spectrum also cannot be reproduced by the secondary electrons from 
the hadronic interaction between the cosmic ray protons and the 
ambient MCs.
Furthermore, the spectral index of the parent electrons of the 
non-thermal X-rays ($s=1.9$--3.4), is larger than the index of the 
electrons generating the synchrotron radio emission ($s<1.7$), 
indicating that the X-ray and radio emission do not share the same 
population of electrons and possibly originate from different shocks.

\item The gas inside the remnant yields a low elemental abundance with 
no evidence of ejecta.
Following a careful background subtraction that minimizes contamination
by the Galactic ridge emission (an improvement over previous X-ray
studies), we find that the spectra of the hot central gas can be best 
fitted with a two-temperature model (\vnei+\vmekal) and also 
can be adequately described with a recombining plasma model (\neij).

\item 
In the two-temperature model (for the central gas), the cold component is 
under-ionized at a temperature $\sim 0.4~\keV$, and the hot component 
is a CIE plasma at $\sim 0.8~\keV$.  
The spatially resolved spectral study reveals variations in temperature, 
gas density, and intervening absorption inside the remnant. 
In the north and east (``CN" and ``C4") of the X-ray-brightness peak, 
the plasma gets relatively cold and dense, and the X-rays suffer heavier 
foreground absorption than those in other parts of the central region,
and filamentary \Ha\ structures are detected. 
This distribution is consistent with the scenario that the remnant is 
evolving in an non-uniform medium with dense gas located in the 
northeast.

\item 
The hot central gas can alternatively be described with a 
multi-$T_z$ recombining plasma model with an electron temperature 
$\sim 0.6~\keV$.
The recombination age of the gas is estimated as $\sim 2.9 \times 
10^4$~yr, slightly lower but close to the dynamical age of \snr\
(3--4$\E{4}$~yr).
The parameters of the recombination model show a slight variation 
among the small regions in the SNR interior, except that the 
northern region ``CN", spatially coincident with the inner radio 
shell, shows the largest recombination timescale and lowest 
ionization temperature.

\item 
The clumpy structure of the central hot gas revealed by the X-ray 
mapping favors the cloudlet evaporation process, which can explain 
the two-temperature components, the low ionization timescale of the 
cold component, and the larger \Ha\ luminosity than that of the X-ray. 
Though, in the two-temperature model, thermal conduction may also 
play a role in a scale as large as the remnant's radius.
{\it ASTRO-H} will help confirm the physical mechanism at play in this, 
and other, MM~SNRs.

\end{enumerate}

{\it Note added in manuscript.} While this paper was under revision, 
Nakamura \etal\ (2014) posted on the arXiv a paper studying the NE 
shell of \snr\ with \XMMN\ observations.
There are some differences between our results and theirs.
We conclude that a two-component is required for the NE shell but 
their fit with a single thermal component with $kT\sim 0.3$~keV is
satisfactory.
The possible discrepancy between our results may be due to the different 
background selection.
They selected a background region inside the SNR where X-ray emission 
is clearly seen (see their Figure 3(a)).
We note that an additional hard component was previously 
pointed out by Ueno et al. (2003), which is consistent with
our finding.

\begin{acknowledgements}
We gratefully acknowledge the anonymous referee for valuable comments.
We are thankful to Ken Tatematsu for providing us with 
the calibrated JCMT CO data.
We also thank Hui Li and Xin Zhou for helpful discussions.
We acknowledge the support from the Natural Sciences and Engineering 
Research Council of Canada (NSERC), the 973 Program grant 2009CB824800, 
the NSFC grant 11233001, the Chinese Scholarship Council,
the grant 20120091110048 from the Educational Ministry of China,
the Canada Research Chairs program and
the Canadian Space Agency.

\end{acknowledgements}

\clearpage



\end{document}